\documentclass{aa}  

\usepackage{graphicx}
\usepackage{txfonts}
\usepackage{subcaption}
\usepackage{amstext}

\begin{document}

   \title{Photoevaporation of the Jovian circumplanetary disk}

   \subtitle{I. Explaining the orbit of Callisto and the lack of outer regular satellites}

   \author{N. Oberg
          \inst{1,2}
          \and
          I. Kamp 
          \inst{1} 
          \and
          S. Cazaux
          \inst{2,3}
          \and
          Ch. Rab
          \inst{1}
          }

   \institute{Kapteyn Astronomical Institute, University of Groningen, P.O. Box 800, 9700 AV Groningen, The Netherlands \\
              \email{oberg@astro.rug.nl}
         \and
             Faculty of Aerospace Engineering, Delft University of Technology, Delft, The Netherlands 
         \and
             University of Leiden, P.O. Box 9513, 2300 RA, Leiden, The Netherlands \\    
             }

   \date{Received --- accepted ---}

 
  \abstract
   { The Galilean satellites are thought to have formed from a circumplanetary disk (CPD) surrounding Jupiter.  When it reached a critical mass, Jupiter opened an annular gap in the solar protoplanetary disk (PPD) that might have exposed the CPD to radiation from the young Sun or from the stellar cluster in which the Solar System formed.}
   {We investigate the radiation field to which the Jovian CPD was exposed during the process of satellite formation. The resulting photoevaporation of the CPD is studied in this context to constrain possible formation scenarios for the Galilean satellites and explain architectural features of the Galilean system.}
   {We constructed a model for the stellar birth cluster to determine the intracluster far-ultraviolet (FUV) radiation field.   We employed analytical annular gap profiles informed by hydrodynamical simulations to investigate a range of plausible geometries for the Jovian gap. We used the radiation thermochemical code \textsc{ProDiMo} to evaluate the incident radiation field in the Jovian gap and the photoevaporation of an embedded 2D axisymmetric CPD. }
   {We derive the time-dependent intracluster FUV radiation field for the solar birth cluster over 10 Myr.  We find that intracluster photoevaporation can cause significant truncation of the Jovian CPD. We determine steady-state truncation radii for possible CPDs, finding that the outer radius is proportional to the accretion rate $\dot{M}^{0.4}$. For CPD accretion rates $\dot M < 10^{-12} M_{\odot}$ yr$^{-1}$, photoevaporative truncation explains the lack of additional satellites outside the orbit of Callisto. For CPDs of mass \mbox{$M_{\rm CPD} < 10^{-6.2} M_{\odot}$} , photoevaporation can disperse the disk before Callisto is able to migrate into the Laplace resonance. This explains why Callisto is the only massive satellite that is excluded from the resonance. }
   {}


   \maketitle
%

\section{Introduction}

We consider the question whether the Galilean moon system is representative of satellite systems of extrasolar giant planets.  The Galilean satellites were formed in a Jovian circumplanetary disk (CPD) \citep{1982Icar...52...14L, 2002AJ....124.3404C, 2003Icar..163..198M} near the end of the formation of Jupiter \citep{2018MNRAS.480.4355C}. While direct detection of extrasolar Galilean analogs has thus far been unsuccessful \citep{2018AJ....155...36T}, several candidate moon-forming CPDs have been identified. The most robust CPD detections are associated with the PDS 70 system; the $5.4\pm1$ Myr old system contains two accreting planets at 23 and 35 au within a cavity  \citep{2018ApJ...863L...8W,2018A&A...617L...2M,2019NatAs...3..749H}.  The inner planet PDS 70b has been detected in near-infrared photometry with an inferred mass of 5-9 $M_{\rm J}$ \citep{2018A&A...617A..44K} and a derived upper limit on circumplanetary dust mass lower than $\sim$0.01$M_{\oplus}$ \citep{2019A&A...625A.118K}.  K-band observations with VLT/SINFONI found a planetary spectral energy distribution (SED) best explained by the presence of a CPD \citep{2019ApJ...877L..33C}. Observations with ALMA at 855 $\mu$m found continuum emission associated with a CPD around PDS 70c, with a dust mass $2\times10^{-3}  - 4.2\times10^{-3}  M_{\oplus}$ and an additional submillimeter point source spatially coincident but still offset from PDS 70b \citep{2019ApJ...879L..25I}.  Additional CPD candidates have been identified in a wide-orbit around CS Cha \citep{2018A&A...616A..79G}, around MWC 758 \citep{2018A&A...611A..74R}, and in the inner cavity of the transitional disk of HD 169142 \citep{2014ApJ...792L..23R}. Because potentially habitable exomoons may outnumber small terrestrial worlds in their respective habitable zones, the prevalence of massive satellites is a question of pertinent astrobiological interest  \citep{2014AsBio..14..798H}.

The core-accretion model suggests that when the core of Jupiter reached a mass of 5-20 $M_{\oplus}$ it began a process of runaway gas accretion \citep{POLLACK199662,2013A&A...558A.113M}, requiring that it formed prior to the dispersal of the gas component of the protoplanetary disk and thus within $\sim$10 Myr of the formation of the Solar System \citep{2001ApJ...553L.153H}.  Gravitational interaction between Jupiter and the circumstellar disk possibly resulted in the rapid opening of an annular gap  \citep{1986ApJ...309..846L,1993prpl.conf..749L,2007MNRAS.381.1280E,2010ApJ...714.1052S, 2012A&A...546A..18M} in which the surface density was reduced relative to the surrounding protoplanetary disk (PPD) by a factor $\sim10^2-10^4$ \citep{1999MNRAS.303..696K, 2017ApJ...842..103S}.   The timescale of the gap opening has been constrained by isotopic analysis of iron meteorites, which suggests that two distinct nebular reservoirs existed within the solar PPD, where the Jupiter gap acted to partially isolate the two reservoirs \citep{2014A&A...572A..35L, Kruijer}.  In this case, the Jupiter core reached a mass of 20 $M_{\oplus}$ within <1 Myr and then grew to 50 $M_{\oplus}$ over 3-4 Myr, which is more slowly than predicted in the classical core-accretion scenario \citep{Kruijer}. 

The accretion of gas onto Jupiter most likely led to the formation of a circumplanetary disk \citep{2008ApJ...685.1220M}.  The precise characteristics of the circumplanetary disk are still unclear, and several competing models are still considered. One possibility is a massive ($\sim 10^{-5} M_{\odot}$) static disk of low viscosity, which either initially contained or was later enriched by sufficient solid material to form the Galilean satellites \citep{1982Icar...52...14L,2003Icar..163..198M, 2018MNRAS.475.1347M}. Alternatively, a family of accretion disk models has been postulated, in which the disks were fed by the continuous inflow of material from the surrounding PPD \citep{2002AJ....124.3404C,2005A&A...434..343A,2006Natur.441..834C,2009euro.book...59C}. Even after the formation of a low-density annular gap, PPD material is expected to continue to flow across the gap and onto the planet and its CPD \citep{2006ApJ...641..526L}. Population synthesis models suggest that an accretion disk can successfully produce a Galilean-like retinue of satellites \citep{2010ApJ...714.1052S, 2018MNRAS.480.4355C}. After an optically thin gap is opened, solar photons may scatter onto the CPD or may impinge directly from interstellar space. 

When a gaseous disk is exposed to an external UV radiation field as is present within a stellar cluster, a thin surface layer of gas can be heated such that the local sound speed of gas exceeds the gravitational escape velocity and the gas becomes unbound. This launches a thermal wind of escaping gas and entrained dust from the disk surface and outer edge \citep{1994ApJ...428..654H, 2000ApJ...539..258R}.  If the mass-loss rate caused by the photoevaporative flow is greater than the radial mass transport by viscous evolution, the disk can become truncated \citep{2007MNRAS.376.1350C}. Photons of energy $6 < h\nu < 13.6$ eV are known as far-ultraviolet (FUV) photons and are expected to dominate  photoevaporation in moderately sized clusters \citep{2004ApJ...611..360A}.  

If the exposed Jovian CPD is depleted of volatiles or disrupted by photoevaporative processes prior to the formation of the Galilean satellites, it must afterward be enhanced in volatiles by a dust- or planetesimal-capturing process, or by continued mass accretion across the gap to explain the satellite compositions \citep{1999ApJ...526.1001L,2006Natur.441..834C, 2010SSRv..153..431M, 2018AJ....155..224R}. While it is possible that satellite formation terminated prior to the stage of the Jovian gap opening  \citep{2010ApJ...714.1052S}, we consider models where the formation of the (final generation of) satellites occurs after the gap opening, in the case of either an optically thick massive and static disk or of a slow-inflow accretion disk \citep{1999ApJ...526.1001L, 1999MNRAS.303..696K, 2002AJ....124.3404C, 2018AJ....155..224R}.  

Several lines of evidence suggest that the Sun was formed in a stellar cluster \citep{2010ARA&A..48...47A} with a virial radius \mbox{$r_{\rm vir} = 0.75\pm0.25$ pc} and number of stars $N = 2500\pm300$  \citep{2019A&A...622A..69P}. The short-lived radio-isotope (SLR) $^{26}$Al in meteorites may have been produced by the enrichment of the solar PPD by winds from a nearby massive Wolf-Rayet star \citep{doi:10.1029/GL003i002p00109,2012E&PSL.359..248T, 2019A&A...622A..69P}.  The truncation of the mass distribution in the Solar System beyond \mbox{45 au} may have been caused by close stellar encounters in the birth cluster, UV-driven photoevaporation by nearby massive stars, or ram-pressure stripping by a supernova blast wave \citep{2004ApJ...611..360A,2009ApJ...696L..13P,2010ARA&A..48...47A}. The cluster was likely an OB association in which a close stellar encounter occurred within 2 Myr and the probability of further close encounters became negligible after 5 Myr \citep{2013A&A...549A..82P}. The Solar System was therefore likely  bathed in an intense FUV radiation field while Jupiter was forming.  The cluster would eventually have dispersed within some 10 to 100 million years \citep{2001ApJ...562..852H}.

The radiation environment inside the gap and around the circumjovian disk has been studied previously. \citet{2012ApJ...748...92T} used a Monte Carlo radiative transfer method to study the intragap radiation produced by a 10 $L_{\odot}$ star at time $t$ = 0.3 Myr, motivated by the very rapid gap opening of a Jupiter formed in the gravitational instability scenario.  Hydrostatic disk flaring in the gap exterior results in an illuminated outer edge of the gap that absorbs stellar radiation and reradiates it into gap, resulting in a hot (> 150 K) gap that is inconsistent with an early formation of satellites \citep{2007prpl.conf..607D}.  Photoevaporation of the circumjovian disk has also been considered analytically in the context of a fixed CPD surface temperature  \citep{2011AJ....142..168M}. In a 1D simulation that considered viscous evolution, accretion, and photoevaporation of the CPD, \citet{2011AJ....142..168M} found that the CPD is radiatively truncated to a fraction of the Hill radius  0.16 $r_{\rm H}$, in contrast to tidal forces, which have been suggested to truncate the CPD to 0.4 $r_{\rm H}$ \citep{2011MNRAS.413.1447M}.   The young Sun had an excess X-ray and UV flux $10^2-10^4$ times greater than at the present day \citep{1982RvGSP..20..280Z, 2002ApJ...572..335F, 2005ApJS..160..401P, 2005ApJ...622..680R}. A rapid migration of Jupiter and its disk to 1.5 au during a possible `Grand Tack' scenario \citep{2011Natur.475..206W} results in a sufficiently high solar irradiation to sublimate the CPD volatile reservoir, implying that the formation of Ganymede and Callisto occurred prior to any inward migration \citep{2015A&A...579L...4H}.  

Several characteristics of the Galilean satellite system remain to be explained.  Callisto is the only Galilean satellite that does not lie within the Laplace resonance, and no additional regular satellites exist beyond the orbit of Callisto.  This might be indicative of an event that abruptly removed the gas content of the CPD to prevent further gas-driven migration by Callisto and to prevent the in situ formation of additional satellites at radii beyond 30 $R_{\rm J}$.  Previously, a rapid dispersal of the Jovian CPD has been invoked to explain the difference in the architectures between the Jovian and Saturnian regular satellite systems \citep{2010ApJ...714.1052S}.  We aim to improve our understanding of the formation of the Galilean satellites by modeling CPD truncation in the presence of external, and in particular, intracluster, FUV radiation.  As a first application, we explore in this paper the conditions under which photoevaporation might explain the lack of massive satellites outside the orbit of Callisto and how it might have prevented Callisto from entering the Laplace resonance.


\section{Methods}

\begin{figure}
  \includegraphics[width=\textwidth/2]{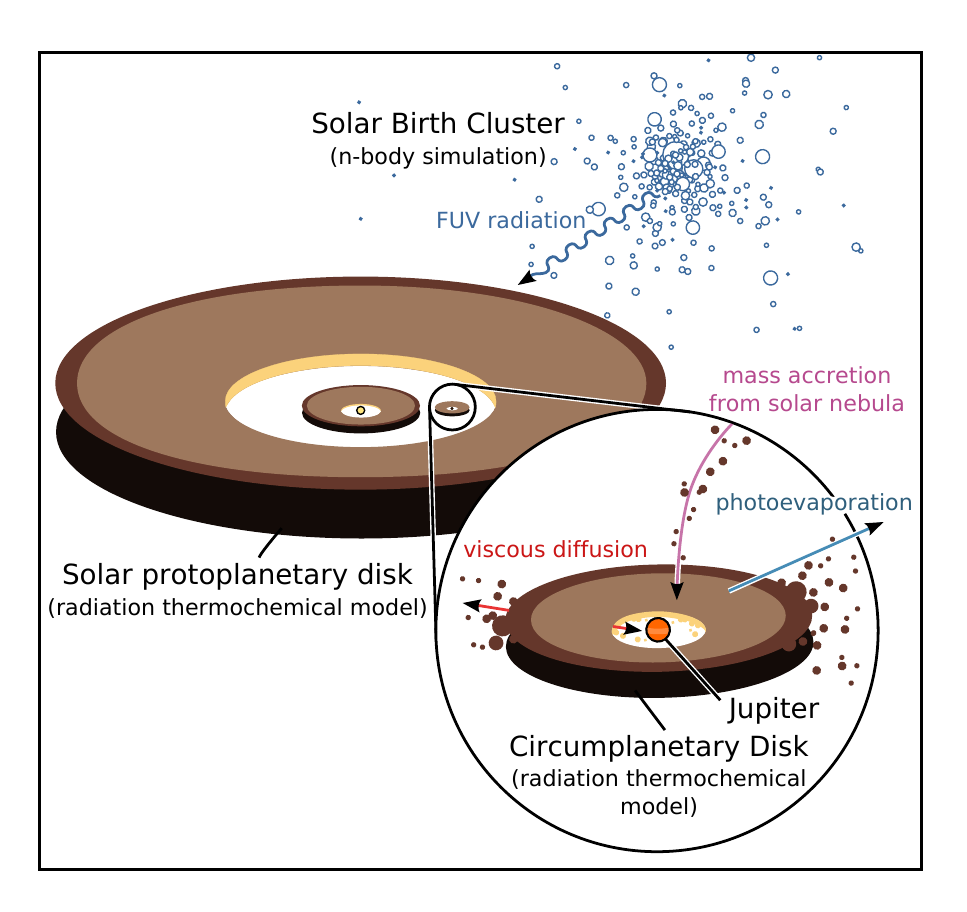}
      \caption{Illustration of the systems we modeled.  The solar birth cluster n-body provides the background radiation field for the solar protoplanetary disk in the model.  The protoplanetary disk model is used to determine the fraction of radiation incident on the circumplanetary disk.   Mass in- and outflow to the CPD is calculated to derive steady-state truncation radii.}
      \label{fig:cartoon}
\end{figure}

To capture the relevant physics of cluster-driven circumplanetary disk photoevaporation across nine orders of magnitude in spatial scale (from tens of Jovian radii to several parsecs) and four orders of magnitude in temporal resolution (from 1 kyr to 10 Myr) we model the evolution of the stellar cluster, Solar System, and Jovian circumplanetary disk independently. The general method of this approach is thus outlined in three parts.  A diagram illustrating the systems modeled is shown in Fig. \ref{fig:cartoon}.  \\
\\
1) In Sect. \ref{sec:method_cluster} we simulate the evolution of an analog to the Sun-forming stellar cluster to determine the time-varying FUV radiation field to which the Sun-like cluster stars are exposed over a typical PPD lifetime of $\sim$ 10 Myr \citep{2010pdac.book..263P}.  Each cluster star is sampled from a Kroupa initial mass function (IMF) \citep{2002Sci...295...82K} , and a stellar FUV luminosity corresponding to its mass is calculated.  Because the precise location of the Solar System inside the birth cluster is not known, we consider the radiation field incident on all cluster stars.  This allows us to study 2500 Jovian CPD analogs simultaneously and to investigate the resulting distribution of incident FUV radiation fields as a function of time and stellar mass. \\
\\
2) In Sect. \ref{sec:prodimo} we construct a radiation thermochemical disk model of the protoplanetary disk of the Sun and introduce an annular gap of width 1-2 au in the surface density profile to investigate the penetration of the stellar and interstellar FUV radiation of the disk midplane and quantify the intragap radiation field.  \\
\\
3) In Sect. \ref{section:loss} the derived intragap radiation field is then applied as a background to a grid of disk models representing plausible Jovian CPDs.  The resulting gas temperature structure and photoevaporative mass loss of the CPD are studied as a function of time.  The truncation radii of the CPDs are calculated, and the evolution of the outer radius as a function of decreasing mass accretion is studied.\\
\\
\noindent

\subsection{Interstellar radiation field and cluster environment} \label{sec:method_cluster}

\begin{figure}
  \includegraphics[width=\textwidth/2]{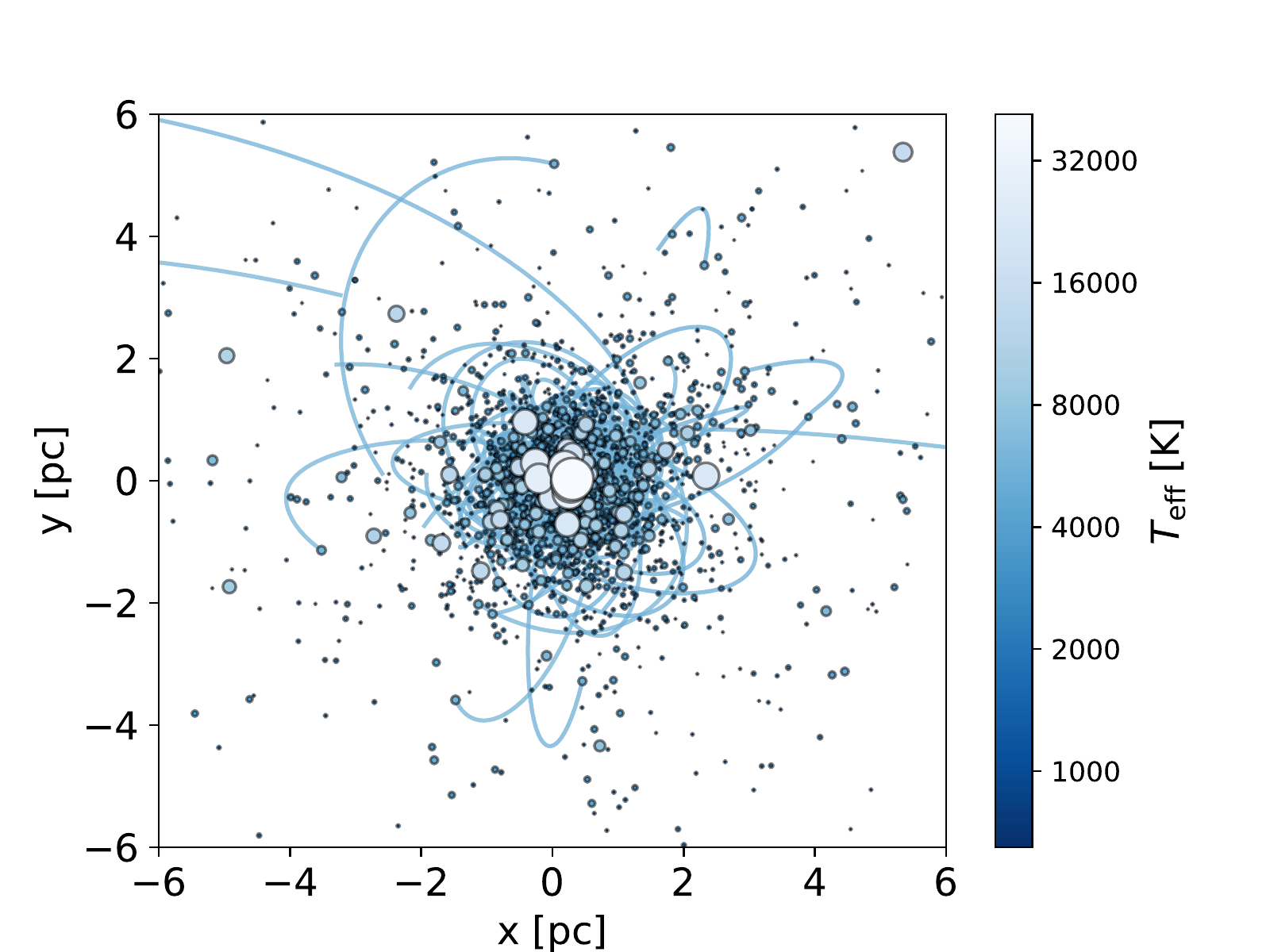}
      \caption{Projected x-y plane distribution of stars in the solar birth cluster at $t = 5$ Myr.  Marker size and color indicate stellar mass. Trajectories of stars with mass corresponding to G-type stars are traced for 5 Myr.}
      \label{fig:clusterxy}
\end{figure}

To determine the UV radiation field that an arbitrary cluster star is exposed to, we created a model of the solar birth environment. For the mass distribution of the cluster, we adopted a simple Plummer sphere \citep{1911MNRAS..71..460P} and initialized it into virial equilibrium.  The cluster was given a number of stars $N$ = 2500 and a virial radius 0.75 pc \citep{2019A&A...622A..69P}.  The spatial distribution, mass, and temperature of the cluster stars are shown in Fig. \ref{fig:clusterxy}.  We sampled the stellar masses from the Kroupa IMF \citep{2002Sci...295...82K} with a lower mass limit of 0.08 $M_{\odot}$ and an upper mass of 150 $M_{\odot}$ \citep{2006MNRAS.365.1333W}.   An arbitrary resampling of 2500 stars from the Kroupa IMF resulted on average in 2.9$\pm$1.5 stars of mass greater than 20 $M_{\odot}$ in the cluster that dominate the cluster FUV emission.  The resulting cluster mass is typically $\sim$ 1000 $M_{\odot}$.  The cluster simulation was then integrated over \mbox{10 Myr} using the REBOUND n-body code \citep{2012A&A...537A.128R}. 

In our disk modeling code \textsc{ProDiMo} (described in Sect. \ref{sec:prodimo}), the intensity of the external UV radiation field is parameterized by the so-called unit Draine field, $\chi$, which is defined as

\begin{equation}
    \chi = \frac{\int_{205 \rm nm}^{91.2 \textrm{nm}} \lambda u_{\lambda} d\lambda}{ \int_{205\textrm{nm}}^{91.2\textrm{nm}} \lambda u^{\textrm{Draine}}_{ \lambda}d\lambda}
,\end{equation}

\noindent
where $ \lambda u_{\lambda}$ is the spectral flux density, and $ \lambda u^{\rm Draine}_{\lambda}$ is the UV spectral flux density of the ISM \citep{1978ApJS...36..595D,2007A&A...467..187R,2009A&A...501..383W}.  For $\chi$ = 1, the  dimensionless Habing unit of energy density $G_0 \approx 1.71$ , and when it is integrated from 6-13.6 eV, $G_0$ = 1 corresponds to a UV flux \mbox{$F_{\rm UV} = 1.6\times 10^{-3}$ erg s$^{-1}$ cm$^{-2}$} \citep{1996ApJ...468..269D}. In the radiative transfer, the external FUV radiation field is approximated as the sum of an isotropic diluted blackbody (20000 K) and the cosmic microwave backgroun (CMB) (2.7 K),

\begin{equation}
    I_{\nu} = \chi 1.71 W_{\rm dil} B_{\nu}(20000K) + B_{\nu}(2.7 K),
    \label{eq:chi}
\end{equation}

\noindent
where $W_{\rm dil}$ = 9.85357$\times10^{-17}$ is the normalization factor for \mbox{$\chi = 1$} \citep{2009A&A...501..383W}. The integration boundaries bracket radiation intensities that drive important photoionization and photodissociation processes in protoplanetary disks \citep{B517564J}. The upper integral boundary of 91.2 nm is the Lyman limit.  In \textsc{ProDiMo,} $\chi$ is a free parameter. We can thus investigate the exposure of the CPD to variable FUV flux levels derived from the stellar cluster model.

Each star in the cluster was taken to be an ideal blackbody with a luminosity determined by the mass-luminosity relation $L \propto M^{3.5}$. We derived the stellar temperature and computed the resulting Planck function \citep{salaris2005evolution}.   We integrated the Planck function over the FUV wavelength range specified in Eq. \ref{eq:chi} to determine the photospheric FUV luminosity per star. T Tauri stars exhibit FUV excess in part because of strong Ly$\alpha$ \mbox{(10.2 eV)} emission, which can contribute $\sim90\%$ of the FUV flux \citep{2005ApJ...622..680R, 2012ApJ...756L..23S}.  We therefore applied a correction factor to our T Tauri cluster stars ($M_* < 8$ $M_{\odot}$) by scaling the UV luminosity relative to the bolometric luminosity $L_{\rm bol}$ to satisfy $L_{\rm UV} \sim 10^{-2.75 \pm 0.65 } L_{\rm bol}$ for classical T Tauris \citep{2012ApJ...744..121Y}.  To reflect the large scatter in this relation, we drew 2500 samples from a normal distribution with standard deviation of the quoted uncertainty and applied to the cluster stars.  The combined effect over the cluster is a FUV luminosity increase by a factor $\sim 60$.  

We then calculated for each star within the cluster the resulting incident FUV flux that originated from all others cluster stars for each time step.  The behavior of the photoevaporative mass-loss rates described in  Sect. \ref{section:loss} was then determined as a function of the time-variable FUV intra-cluster radiation field.  We also considered that the CPDs are partially shielded by the surrounding disk material exterior to the gap and that the interstellar FUV flux must be reduced by a fraction corresponding to this opening angle.  We here considered six surface density profiles to represent the disk evolution as a function of time. The profiles are shown in Fig. \ref{fig:sdps}.  

Two phenomena act to reduce flux incident on the CPD.  The first is shielding from the surrounding protoplanetary disk material external to the gap.  The second is the orientation of the system relative to the primary source of FUV radiation, which are the B stars in the cluster central region. Because we did not monitor the inclination of the PPDs during the cluster simulation, we averaged over all possible disk inclinations to account for the projected area and shielding from the PPD.

For each profile we measured the opening angle over which the planet has an optically thin line of sight to the exterior.  We find visual extinction A$_{\rm V} < 1$ opening angles for the six profiles in Fig. \ref{fig:sdps}. These correspond to shielding of isotropic incident flux by a factor 0.5, 0.33, and 0.2 for \mbox{$t = 1$ Myr}, \mbox{$\alpha = 10^{-3}, 10^{-4}, \text{and } 10^{-5}$} , respectively, and 0.12, 0.06, and 0.03 for \mbox{$t = 5$ Myr}, \mbox{$\alpha = 10^{-3}, 10^{-4},\text{and } 10^{-5}$} , respectively.   We smoothly interpolated this shielding factor from the maximum to the minimum for the assumed $\alpha$-viscosities over the first 5 Myr of the simulation to represent the evolution of the disk surface density. The effective projected area of a disk averaged across all possible orientations in three dimensions is half of its true area.  At $t = 1$ Myr, the flux incident on the CPDs is thus diluted by a factor 4 at most, independently of extinction of normally incident rays.  When instantaneous snapshots of the FUV flux distribution were extracted (as in Fig. \ref{fig:t_g0}), we convolved a random distribution of disk inclinations with the flux distribution to compensate for the effect of the averaged incidence angle.

Because the cluster includes short-lived massive (\mbox{$M > 20$ $M_{\odot}$}) stars that contribute to the total cluster FUV luminosity, it is worthwhile to discuss the possibility and implication of supernovae explosions that might occur within the cluster. While a supernova in the vicinity of the Sun may be required to explain r-process anomalies in Ca-Al inclusions, such an event would have occurred during the earliest formation stages of the Solar System \citep{Brennecka17241} and it is therefore not relevant for the satellite formation process of Jupiter.  However, the lifetime of massive stars greater than \mbox{20 $M_{\odot}$} is \mbox{$\leq$ 8 Myr} \citep{1992A&AS...96..269S}, and it is therefore possible that other stars ended their main-sequence lifetime during the 10 Myr time span considered in our simulation.  We find that when we populate our model cluster by sampling stellar masses from the Kroupa IMF, the population of stars of \mbox{$M > 20$ $M_{\odot}$} typically contribute $\sim75\%$ of the integrated cluster FUV luminosity, with the individual massive stars contributing $\sim25\%$ each to the total cluster FUV.  The effect of the satellite formation process is thus contingent on whether the most massive stars formed early or late in the stellar formation history of the cluster. In clusters of $10^3-10^4$ members, the majority of star formation is expected to occur contemporaneously; in Orion, it is estimated that 53$\%$ of cluster stars formed within the last \mbox{1} Myr and $97\%$ within the last 5 Myr \citep{2000ApJ...540..255P}.  The apparent ejection of several high-mass stars from Orion \mbox{2.5 Myr} ago and the current ongoing massive star formation implies that both early and late formation of massive stars are possible \citep{2001A&A...365...49H, 2007prpl.conf..165B}. We expect satellite formation to have occurred no earlier than 4 Myr after the formation of the Ca-Al inclusions to satisfy the constraint of the internal differentiation state  of Callisto and no later than the dispersal of the gas component of the solar PPD after \mbox{$\sim 5$ Myr} \citep{2002AJ....124.3404C,2008Icar..198..163B, 2009AIPC.1158....3M}.  If the massive star formation rate is independent of the cluster age and the overall star formation rate accelerates, as observed in Orion, there is a 90-95$\%$ probability that a 20 $M_{\odot}$ star will not explode prior to the stage of satellite formation.

In the eventual case of one or more supernovae, we expect an abrupt reduction in cluster FUV luminosity by 25-75$\%$. By resampling from our adopted IMF and reinitializing the initial positions of cluster members, we find that a cluster with $70\%$ lower FUV flux than our fiducial case is still 1$\sigma$ consistent with the resulting spread in cluster FUV luminosities. While the Solar System might be ram-pressure stripped by supernovae explosions \citep{2018A&A...616A..85P}, the CPD is an actively supplied accretion disk whose mass and steady-state radius depend on the accretion rate, and we therefore expect CPDs to rebound to steady state within some $10^3$ yr after accretion is resumed.

\subsection{Disk modeling with \textsc{ProDiMo}} \label{sec:prodimo}

We used the radiation thermochemical disk model \textsc{ProDiMo} \footnote{https://www.astro.rug.nl/~prodimo/} (protoplanetary disk model) to simulate the solar nebula at the stage before the gap opening of Jupiter \citep{2009A&A...501..383W,2010A&A...510A..18K,2011MNRAS.412..711T}.  \textsc{ProDiMo} calculates the thermochemical structure of disks using a frequency-dependent 2D dust continuum radiative transfer, gas-phase and photochemistry, ice formation, and nonlocal thermal equilibrium (NLTE) heating and cooling mechanisms to self-consistently and iteratively determine the physical, chemical, and radiative conditions within a disk.  \textsc{ProDiMo} performs a full 2D ray-based wavelength-dependent continuum radiative transfer at every grid point in the disk to calculate the local continuous radiation field $J_{\nu}(r,z)$ \citep{2009A&A...501..383W}.  Rays are traced backward from each grid point along their direction of propagation while the radiative transfer equation is solved assuming LTE and coherent isotropic scattering.   For a full description of the radiative transfer method, see \citet{2009A&A...501..383W}.  We adopted the standard DIANA dust opacities \footnote{https://dianaproject.wp.st-andrews.ac.uk/data-results-downloads/fortran-package/} for a mixture of amorphous silicates, amorphous carbon, and vacuum  \citep{2016A&A...586A.103W,2016A&A...585A..13M}.

\subsubsection{Solar protoplanetary disk}

\begin{figure}
  \includegraphics[width=\textwidth/2]{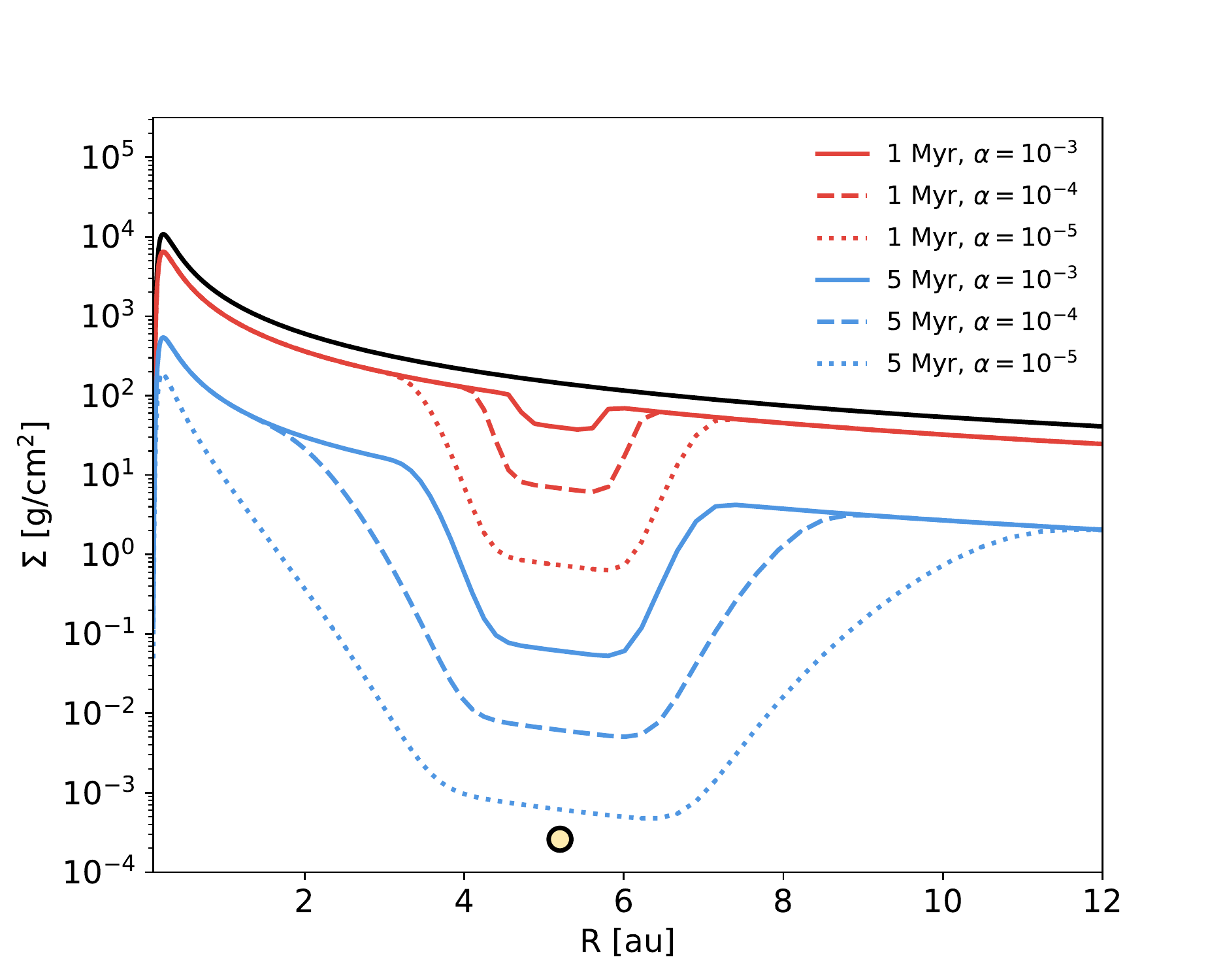}
      \caption{Surface density profiles of the solar protoplanetary disk.  The red profiles show the earliest gap-opening phase when \mbox{$M_{\rm p}$ = 0.1 $M_{\rm J}$} case at $t = 1 $Myr when the disk has been reduced to 60$\%$ of its initial mass.  The blue profiles show the \mbox{$M_{\rm p}$ = $1 M_{\rm J}$} case where the remaining disk mass is $5\%$ of the initial mass.  The black line plots the unperturbed disk surface density at $t_0$ with a total mass of 0.02$M_{\odot}$.  The orange dot indicates the radial location of Jupiter. }
      \label{fig:sdps}
\end{figure}

To construct our protosolar nebula, we used a modified Hayashi minimum mass Solar nebula (MMSN) surface density profile \citep{1981PThPS..70...35H} combined with an analytical gap structure approximation based on a generalized normal (Subbotin) distribution, with a characteristic flat bottom and Gaussian wings \citep{Sub23}.  The unperturbed surface density $\Sigma$ at a radius $r$ is thus described by

\begin{equation}
 \Sigma(r) =  \Sigma_{\rm 1au} \bigg(\frac{r}{1 \: \textrm{au}}\bigg)^{-3/2} \textrm{ g cm}^{-2},
\end{equation}
\noindent
where  $\Sigma_{\rm 1au}$ is the surface density at $r$ = 1 au, and the gap structure $G$ at a radius $r$ is defined by

\begin{equation}
 G(r) =    X   \frac{ a }{2 b  \Gamma(\frac{1}{a})} \textrm{exp}\bigg(-\bigg|\frac{r-r_{\rm p}}{b}\bigg|^{a}\bigg) \: ,
\end{equation}
\noindent
where $b$ is the standard deviation of the Gaussian component of the gap, $r_{\rm p}$ is the radial location of the gap-opening planet, $a$ is a shape parameter, $\Gamma$ is the gamma function, and the gap depth is controlled by a normalization function $X$ that scales the gap depth relative to the unperturbed surface density profile. The final perturbed surface density profile is thus defined as \mbox{$\Sigma$(r) $\cdot G$(r)}, with a total mass of 0.02 $M_{\odot}$ out to 100 au for a surface density at 1 au  $\Sigma_{\rm 1 au}$ of 1700 g cm$^{-2}$ and a shape parameter $a = 8$.    The semimajor axis of Jupiter at the stage of gap opening is poorly constrained because Jupiter may have migrated to its present location by gravitational interaction, with the protosolar nebula leading to short periods of dynamical instability \citep{2005Natur.435..459T,2011Natur.475..206W}. To make as few assumptions as possible, we therefore placed Jupiter at its current location of $r_{\rm p}$ = 5.2 au.  Our gap dimensions are informed by the analytical gap-scaling relation derived from hydrodynamical simulations of \citet{2016PASJ...68...43K} for Jupiter-mass planets where the intragap minimum surface density $\Sigma_{\rm gap}$ and unperturbed surface density $\Sigma _0$ are related by $\Sigma_{\rm gap}/  \Sigma_0$ = (1+0.04 $K$)$^{-1}$ , with $K$ defined as

\begin{equation}
    K = \bigg( \frac{M_{\rm p}}{M_{\rm*}}\bigg)^2 \bigg( \frac{H_{\rm p}}{r_{\rm p}}\bigg)^{-5} \alpha^{-1}.
\end{equation}
\noindent
Here $H_{\rm p}$ is the disk scale height at the radial location of the planet $r_{\rm p}$, and $\alpha$ is the viscosity.  To determine $H_{\rm p}$/$r_{\rm p}$ , we first ran a single \textsc{ProDiMo} model of the unperturbed solar nebula.  The heating-cooling balance of the disk was iteratively calculated until the disk structure converged to a vertical hydrostatic equilibrium, from which we extracted the radial scale height profile.  Similarly, we employed the formula for the gap width at half-depth of \citet{2016PASJ...68...43K}, where

\begin{equation}
    \frac{\Delta_{\rm gap}}{r_{\rm p}} = 0.41 \bigg(\frac{M_{\rm p}}{M_{\rm*}}\bigg)^{1/2} \bigg( \frac{H_{\rm p}}{r_{\rm p}}\bigg)^{-5} \alpha^{-1/4},
\end{equation}
\noindent
where $\Delta_{\rm gap}$ is the gap depth.  We adopted two cases for the planet mass $M_{\rm p}$ (0.1 and 1 $M_{\rm J}$) owing to the continuing accretion of Jupiter after the gap opening.  We also considered that the total disk mass was reduced over the course of 5 Myr. Numerical simulations of accreting planets in gaseous disks show that the circumstellar disk is reduced to 60$\%$ of its unperturbed mass by the time of the Jovian gap opening, and to only $5\%$ when the planet has reached its final mass of 1 $M_{\rm J}$ \citep{2010exop.book.....S}. We first considered a gap profile at the time of gap opening \mbox{$t \sim 1$ Myr}  when Jupiter has reached $10\%$ of its final mass, or $\sim 30$ $M_{\oplus}$ \citep{2017ApJ...835..146D}, and second, a gap profile for when Jupiter approaches its final mass at $t \sim 5$ Myr. The resulting surface density profiles are shown in Fig. \ref{fig:sdps}.  Based on an initial Hayashi MMSN mass of $\sim0.02 M_{\odot}$, we adopted total protosolar disk masses of $0.012$ $M_{\odot}$ and 0.001 $M_{\odot}$ for the 1 and 5 Myr stages, respectively \citep{2010exop.book.....S}. These values are in general agreement with the observed exponential decrease of the inner disk fraction with time $e^{-t/\tau_{\rm disk}}$ with a disk e-folding time \mbox{$\tau_{\rm disk}$ = 2.5 Myr}. This suggests gas disk masses of 0.67 and 0.13 of the initial mass at 1 and 5 Myr, respectively \citep{2009AIPC.1158....3M}. Finally, because of the uncertainty in the viscosity, we considered a range of possible PPD $\alpha$ viscosities from $10^{-5} -10^{-3}$  \citep{1973A&A....24..337S,2015ApJ...806L..15K, 2017ApJ...837..163R}. The resulting surface density profiles for all self-consistent combinations of planet mass, disk mass, and disk viscosity are plotted in Fig. \ref{fig:sdps}, and they result in a variation of approximately six orders of magnitude in the intragap surface density.  

Solar luminosity and temperature were selected from the evolutionary tracks from the Grenoble stellar evolution code for pre-main-sequence stars \citep{2000A&A...358..593S}.  For the Sun, we adopted a 1 $M_{\odot}$ star with metallicity Z = 0.02 at ages of 1 Myr and 5 Myr to reflect the stages of gap opening and when the mass of Jupiter reaches $\sim 1$ $M_{\rm J}$.  We used the PHOENIX library of stellar atmospheric spectra \citep{2005ESASP.576..565B, 2013A&A...553A...6H}. All parameters of the solar protoplanetary disk are listed in Table \ref{tab:ppds}.

\begin{table}
    \centering
   \begin{tabular}{lll}
    \hline \hline
        Parameter               & Symbol             & Value  \\ \hline
        Stellar mass            & $M_*$              & 1.0 $M_{\odot}$           \\
        Stellar luminosity      & $L_*$              & 2.335, 0.7032 $L_{\odot}$ \\
        
        Effective temperature   & $T_{\rm eff}$      & 4278, 4245 K                    \\
        UV luminosity           & $L_{\rm UV,*}$       & 0.01 $L_{\odot}$ \\
        X-ray luminosity        & $L_{\rm X}$        & 10$^{30}$ erg s$^{-1}$          \\
        Interstellar UV field   & $\chi$       &  $10^1-10^7$                            \\ 

        \hline

        Disk mass               & $M_{\rm disk}$    & 0.012, 0.001  $M_{\odot}$ \\
        Disk inner radius       & $R_{\rm in} $  & 0.1 au \\ 
        Disk outer radius       & $R_{\rm out} $ & 100 au \\

        \hline
        Minimum dust size & $a_{\rm min}$ & 0.05 $\mu$m\\
        Maximum dust size & $a_{\rm max}$& 3000 $\mu$m \\
        Dust size power law & p & 3.5\\
        Dust-to-gas ratio & $d/g$ & $10^{-2}$\\
        Dust composition: \\
        Mg$_{0.7}$Fe$_{0.3}$SiO$_3$ & & 60 $\%$ \\
        Amorphous carbon & &  $15\%$\\
        Vacuum &  & $25\%$\\

        \hline
        Viscosity & $\alpha$ & $10^{-3},10^{-4},10^{-5}$

    \end{tabular}
    \caption{\textsc{ProDiMo} model parameters for the solar protoplanetary disk at ages $t$ =  1 Myr and 5 Myr.  Stellar temperature and luminosity are selected from the pre-main-sequence stellar evolutionary tracks of \citet{2000A&A...358..593S}. Stellar UV and X-ray luminosities are adopted from \citet{2016A&A...586A.103W}. }
    \label{tab:ppds}
\end{table}

\subsubsection{Circumplanetary disk}

Jovian CPD models can be roughly sorted into two categories.  The first is a CPD analogous to the MMSN that contains sufficient solid matter to construct the Galilean satellites \citep{1982Icar...52...14L,2003Icar..163..198M, 2018MNRAS.475.1347M}. Assuming a canonical dust-to-gas mass ratio of 0.01, the resulting disk has a gas mass of $M_{\rm disk} \approx 2\times10^{-5} M_{\odot}$.  To prevent rapid radial migration and loss of the satellites, the disk must be highly inviscid with $\alpha \leq 10^{-6}$ \citep{2002AJ....124.3404C}. We refer to this class of CPD as the "static" CPD.  The second class of CPD models considers an accretion disk fed by a continuous flow of material from the surrounding circumstellar disk \citep{2002AJ....124.3404C,2005A&A...434..343A,2009euro.book...59C}.  In this class, the instantaneous mass of this CPD is a function of the mass-inflow rate and disk viscosity, and it can be sufficiently low to become optically thin. We therefore consider CPD masses in the range $10^{-5}-10^{-9} M_{\odot}$ \citep{2006Natur.441..834C}.  We refer to this class of CPD as the "accretion" CPD. 

To determine the temperature structure and hence photoevaporation rate of the accretion CPD, we considered the effects of viscous heating through dissipation.  \textsc{ProDiMo} includes the parametrization for viscous heating of \citet{1998ApJ...500..411D}. We specified the viscous heating rate with a mass accretion rate $\dot{M}$. We assumed that the gravitational energy released is converted into heat.  For a steady-state disk, the mass flux $\dot{M}$ is constant between all annuli. The half-column heating rate is 

\begin{equation}
    F_{\rm vis} (r) = \frac{3 G M_{\rm p} \dot{M}}{8 \pi r^3} (1-\sqrt{r_p /r})   \hspace{0.1cm} \textrm{erg cm}^{-2}  \textrm{s}^{-1},
\end{equation}

\noindent
where G is the gravitational constant, $M_{\rm p}$ is the planetary mass, $r$ is the distance to the planet in cylindrical coordinates, and $r_{\rm p}$ is the planetary radius. We must make an assumption about how heat is distributed within the column as a function of height $z$, converted into heating rate per volume by the relation

\begin{equation}
    \Gamma_{\rm vis}(r,z) =  F_{\rm vis} (r) \frac{\rho^P(r,z)}{\int \rho^P(r,z') dz'} \textrm{erg cm}^{-3} \textrm{s}^{-1} ,
\end{equation}
where $\rho$ is the volume density at radial position $r$ and height $z$, and $P$ is a constant equal to 1.5 to avoid unphysical heating at low volume densities.  Accordingly, most of the dissipative heating occurs in the CPD midplane.

\begin{table}
    \centering
   \begin{tabular}{lll}
    \hline \hline
        Parameter               & Symbol             & Value  \\ \hline
        Planet mass             & $M_{\rm p}$           &    1.0 $M_{\rm J}    $      \\
        Planetary luminosity      & $L_{\rm p}$              & $10^{-5} L_{\odot}$ \\
        
        Effective temperature   & $T_{\rm eff,p}$      & 1000 K                    \\
        UV luminosity           & $L_{\rm UV,p}$       & 0.01,0.1 $L_{\rm p}$ \\
        Interstellar UV field   & $\chi$       &  $10^1-10^7$ \\ 
        Background temperature  & $T_{\rm back}$ & 70 K \\

        \hline

        Disk mass               & $M_{\rm cpd}$    & 10$^{-5} - 10^{-9}  M_{\odot} $\\
        Disk inner radius       & $R_{\rm in,cpd} $  & 0.0015 au \\ 
        Disk outer radius       & $R_{\rm out,cpd} $ & 0.2 au \\
        Column density power index & $\beta_{\Sigma}$ & 1.0 \\
        
        \hline
        
        Maximum dust size & $a_{\rm max}$& 10, 3000 $\mu$m \\
        Dust-to-gas ratio & $d/g$ & $10^{-2},10^{-3},10^{-4}$  \\       
        Flaring index & $\beta$ & 1.15 \\
        Reference scale height & $H_{\rm 0.1 au}$ & 0.01 au  \\

        \hline
        
        Accretion rate     &  $\dot M$ & $10^{-12} - 10^{-9} M_{\odot}$ yr$^{-1}$ \\

    \end{tabular}
    \caption{\textsc{ProDiMo} model parameters for models of the Jovian circumplanetary disk. The ranges of CPD mass and accretion rate are subdivided in steps of 10.  Where not specified, the CPD parameters are identical to those listed in Table \ref{tab:ppds}. }
    \label{tab:cpds}
    
\end{table}

The CPD is also heated directly by the radiation of Jupiter.  The early luminosity of Jupiter spans five orders of magnitude over 3 Myr  \citep{2007ApJ...655..541M}.  In our CPD model we considered the Jupiter luminosity $L_{\rm p}$ after the runaway-gas-accretion phase when it briefly peaked at $L_{\rm p} > 10^{-3} L_{\odot}$ and then declined to $\sim 10^{-5} L_{\odot}$ and below.  The surface temperature of Jupiter was likely 500-1000 K at this stage \citep{2002AJ....124.3404C, 2012ApJ...745..174S}.  For the Jovian SED we adopted the DRIFT-PHOENIX model spectra for a body of $t = 1000$ K and surface gravity 2.55$\times10^{3}$ cm s$^{-2}$ \citep{2008ApJ...675L.105H}.

We followed the steady-state surface density formulation of \citet{2002AJ....124.3404C}, which relates $\Sigma \propto \dot{M}/\alpha$. Because we explored five orders of magnitude of CPD mass and three orders of magnitude of mass accretion rate onto the CPD, each combination of CPD surface density $\Sigma$ and accretion rate $\dot M$ corresponds to a unique value of $\alpha$.  Hence we explored CPD $\alpha$ viscosities from $2.5\times10^{-6}-2.5\times10^{-1}$.  Because the mechanism that causes viscosity in the PPD and CPD may not be the same, we considered that the two can have differing $\alpha$-viscosities.  The case of the CPD with a mass of 10$^{-9} M_{\odot}$ and an accretion rate of \mbox{10$^{-9} M_{\odot}$ yr$^{-1}$} is thus rendered unphysical because $\alpha > 1$ is required.  

In selecting a range of dust-to-gas ratios, we considered that dust can either quickly coagulate into satellitesimals through the streaming instability \citep{2018ApJ...866..142D}, or quickly settle into the midplane and be viscously transported into the planet on short timescales.  In either case, the amount of material stored in dust grains is rapidly depleted,  therefore we considered cases in which the dust-to-gas ratio is  $10^{-2}-10^{-4}$ .  Parameters for the static and accretion CPDs are listed in Table \ref{tab:cpds}. 

We considered five CPD steady-state masses in combination with four CPD mass accretion rates for a total of 19 unique CPD models. Furthermore, the five CPD mass models were varied in their dust-to-gas ratio, maximum dust size, and planetary luminosity by the ranges listed in Table \ref{tab:cpds}.

\subsection{Photoevaporative mass loss} \label{section:loss}

Molecular hydrogen in the disk can be heated to $T > 10^4$ K by EUV radiation, but only begins to drive mass loss for massive stellar encounters of $d \ll 0.03$ pc \citep{1998ApJ...499..758J}.  FUV photons penetrate below the ionization front and instead heat the neutral hydrogen layer to $\sim10^3$ K \citep{1998ApJ...499..758J}.  Where the gas heating drives the sound speed of the gas $c_{\rm s}$ above the local escape velocity, the gas becomes unbound, launching a supersonic outward flow of disk material.   The radius beyond  which this occurs is the gravitational radius $r_{\rm g}$, 

\begin{equation}
   r_{\rm g} = \frac{G M_{\rm p} }{c_{\rm s}^2}
,\end{equation}
where $G$ is the gravitational constant, assuming the disk mass \mbox{$M_{\rm disk} \ll M_{\rm p}$}, and where the sound speed is defined as
\begin{equation}
    c_{\rm s} = \sqrt{\frac{\gamma \textrm{k}_B T}{\mu m_H}}.
\end{equation}

\noindent
Here $\gamma$ is the adiabatic index, $T$ is the gas temperature, $k_{\rm b}$ is the Boltzmann constant, $\mu$ is the mean molecular mass, and $m_{\rm H}$ is the mass of hydrogen.  Because \textsc{ProDiMo} locally determines the sound speed at all grid points, we conservatively estimated the resulting mass outflow where $c_{\rm s}$ exceeds the local escape velocity of the Jovian CPD.  At a semimajor axis of 5 au, a 1 $M_{\rm J}$ planet orbiting a 1 $M_{\odot}$ star has a Hill radius $r_{\rm H} \approx 0.34 $ au or 711 $R_{\rm J}$.  Gravitational interaction with the disk can act to perturb the CPD gas and truncate the disk at $\sim 0.4$ $r_{\rm H}$ \citep{2011ESS.....2.3311M}, and the Jovian CPD may therefore not have had a radius larger than $\sim0.14$ au or 290 $R_{\rm J}$.  For a surface layer of the gas disk with a typical speed of sound $c_{\rm s} \sim 5$ km s$^{-1}$ , the associated gravitational radius is 0.03 au. The CPD is thus said to exist in the supercritical regime. The current semimajor axis of the outermost Galilean satellite, Callisto, lies at 0.0126 au, or 0.42 $r_{\rm g}$.  Interestingly, given that gap opening may have occurred as early as when Jupiter grew to \mbox{$M_{\rm p} = $ 0.1 $M_{\rm J}$}, the associated $r_{\rm g}$ lies between the present-day semimajor axes of Ganymede and Callisto at a distance of 0.01 au from Jupiter. 

The mass-loss rate per annulus in the CPD at $r > r_{\rm g}$ can be estimated $\dot{M}_{\rm evap} \approx \rho c_{\rm s}$d$a$ , where $\rho$ is the volume density of gas at the base of the heated layer, and d$a$ is the surface area of the annulus. The escape velocity is $\sqrt{ 2GM_{\rm p}/r}$. To determine the mass-loss rate as a function of $\chi$ , we ran seven \textsc{ProDiMo} models for each of the five considered CPD masses.  Each successive step in the model grid increased the background $\chi$ field by a factor ten, covering the range $10^1-10^7$.  At every radial grid point in the resulting \textsc{ProDiMo} disk models, we determined the lowest height above the midplane $z$ at which the escape criterion was satisfied. The sound speed at the coordinate $c_{\rm s}(r,z,\chi)$ and volume density $\rho(r,z)$ were calculated to determine the mass loss per unit area.  The surface area of the corresponding annulus was then multiplied by the area mass-loss rate to determine a total evaporation rate per annulus, such that   

\begin{equation}
    \dot{M}_{\rm evap}(r,\chi(t)) = c_{\rm s}(r,z,\chi(t)) \rho(r,z) A(r) \: ,
\end{equation}

\noindent
where $A(r)$ is the area of the annulus at radius $r$.  The resulting radial evaporation rate profile was log-interpolated between the seven values of $\chi$, such that any value within the range $10^1-10^7$ could be considered. This is required because the FUV field to which the cluster stars are exposed varies smoothly as a function of time (see Sect. \ref{sec:isrf}).

\subsection{Accretion and viscous evolution of the CPD}

In our model, mass accretion and viscous evolution of the CPD will act to replace mass lost by photoevaporation.  Following from angular momentum considerations, it is believed that matter would accrete not evenly over the CPD surface, but concentrated near a centrifugal radius of $\sim$20 $R_{\rm J}$ \citep{2008ApJ...685.1220M}.  In each global iteration of our CPD photoevaporation model, the accreted material was  distributed in steps radially outward from the centrifugal radius until the steady-state surface density profile was achieved or the accreted mass budget was exhausted.  How rapidly this material can be transported radially is set by the viscous diffusion timescale $ \tau_{\rm visc}$ . A viscously evolving $\alpha$-disk has a global viscous diffusion timescale

\begin{equation}
    \tau_{\rm visc} \approx \frac{r_{\rm CPD}^2}{\nu} = \frac{r_{\rm CPD}^2}{\alpha H^2 \Omega _{\rm k}} ,
\end{equation}

\noindent
where $\nu$ is the kinematic viscosity and $\Omega_{\rm k}$ is the Keplerian angular frequency \citep{1973A&A....24..337S,2009euro.book...59C,armitage2019protoplanetary}.  For $\alpha = 10^{-2}-10^{-5}$ and \mbox{$r_{\rm CPD} = 500$ $R_{\rm J}$},  we find $\tau _{\rm visc} \approx 10^{3}-10^{6}$ yr.  The magnitude of CPD viscosity is highly uncertain.  An $\alpha$ on the order $10^{-3} - 10^{-2}$ allows for Ganymede-sized moons to migrate inward under type I migration and establish the Laplace resonance \citep{2002Sci...298..593P}. Simulations of turbulence induced by magnetorotational instability (MRI) in CPDs suggest values of $\alpha$ lower than $10^{-3}$ \citep{2014ApJ...785..101F}, but baroclinic instabilities have been suggested as means to transport angular momentum in disks \citep{2011A&A...527A.138L}.  

While very high accretion rates of $\dot M \sim 10^{-8} M_{\rm \odot}$ yr$^{-1}$ have been shown to be possible across a gap for a PPD with \mbox{$\alpha = 10^{-3}$} and $H/r = 0.05$ \citep{1999MNRAS.303..696K}, we also considered cases more similar to the slow-inflow accretion disk scenario where \mbox{$\dot M \sim 10^{-11}-10^{-10} M_{\rm \odot}$ yr$^{-1}$} \citep{2002AJ....124.3404C}, which is also consistent with the accretion rate of the PDS 70b CPD candidate  $\dot M \sim 10^{-10.8}-10^{-10.3} M_{\rm \odot}$ yr$^{-1}$ \citep{2019ApJ...877L..33C} and the presence of volatiles in the Jovian CPD midplane \citep{2002AJ....124.3404C}. Hence we considered the range of accretion rates listed in Table \ref{tab:cpds}.

\section{Results}

We simulated the orbital evolution of 2500 stars in the stellar cluster over 10 Myr.  We derived incident intracluster radiation field strengths for all of the cluster stars as a function of time.  To determine the penetration of this radiation to the CPD, the optical depth and background temperature within the Jovian gap were determined.  The heating of the Jovian CPD by the Sun, Jupiter, and intracluster radiation was determined by means of a dust and gas radiation thermochemical model.  The resulting gas properties were used to  simulate the photoevaporation and radiative truncation of circumplanetary accretion disks as a function of time.

\subsection{Interstellar radiation field within the cluster} \label{sec:isrf}

\begin{figure}
  \includegraphics[width=\textwidth/2]{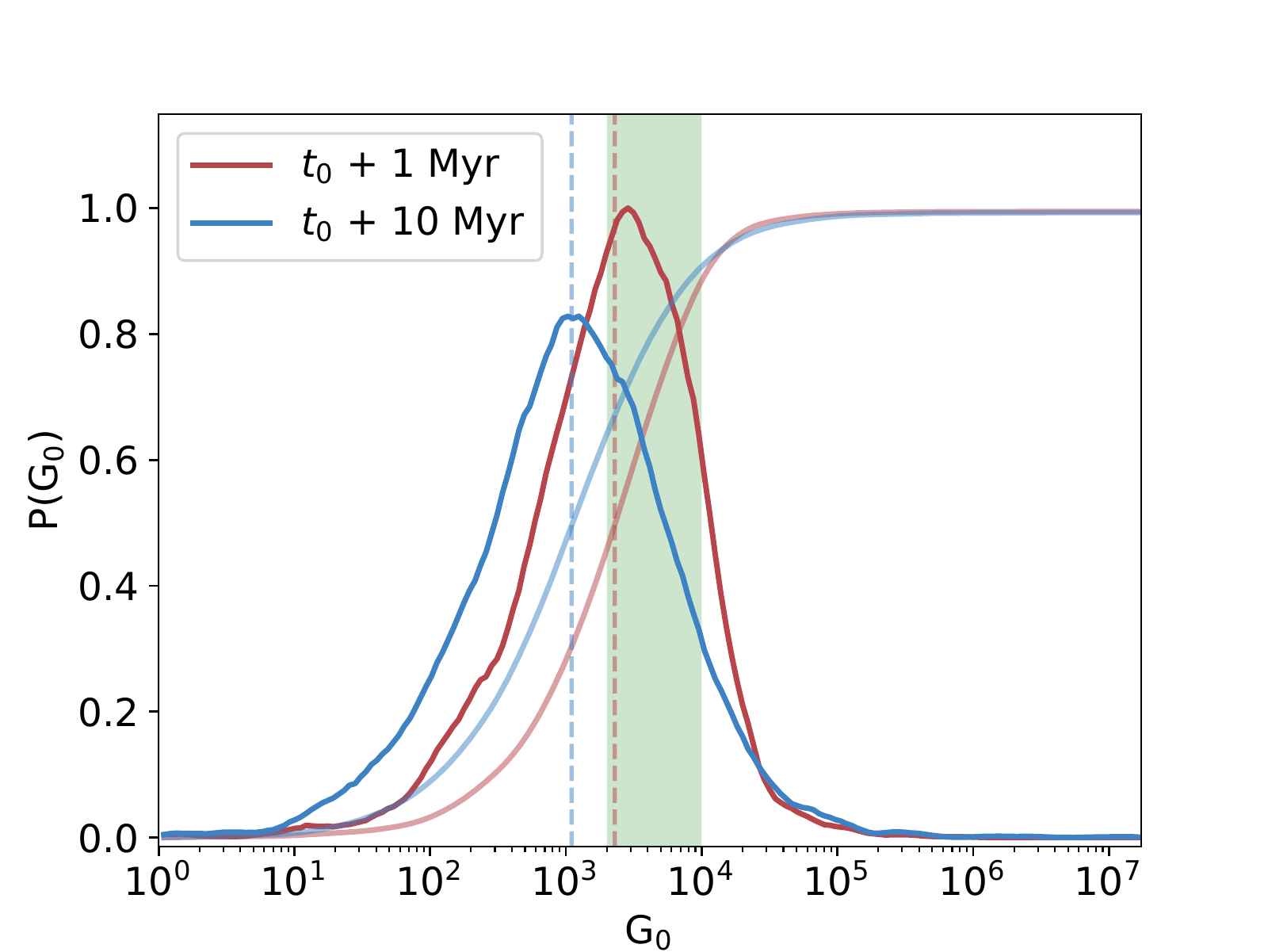}
      \caption{Probability distribution of the intracluster radiation field strength $G_0$ at the position of each cluster star over time, normalized to the maximum likelihood value at $t = t_0 + 1$ Myr.  The distribution evolves over 10 Myr from $t = t_0 + 1$ Myr (red) and to $t = t_0 + 10$ Myr (blue). The faded solid lines illustrate the cumulative distributions.  The vertical dashed lines indicate the median value of each distribution. The green bar indicates the Solar System constraint of \mbox{$2000 \leq G_0 \leq 10^4$}. Multiple initializations of the cluster have been averaged to minimize the Poisson noise because the number of stars is small ($N_{\rm stars} = 2500$).  }
      \label{fig:t_g0}
\end{figure}

\begin{figure}
  \includegraphics[width=\textwidth/2]{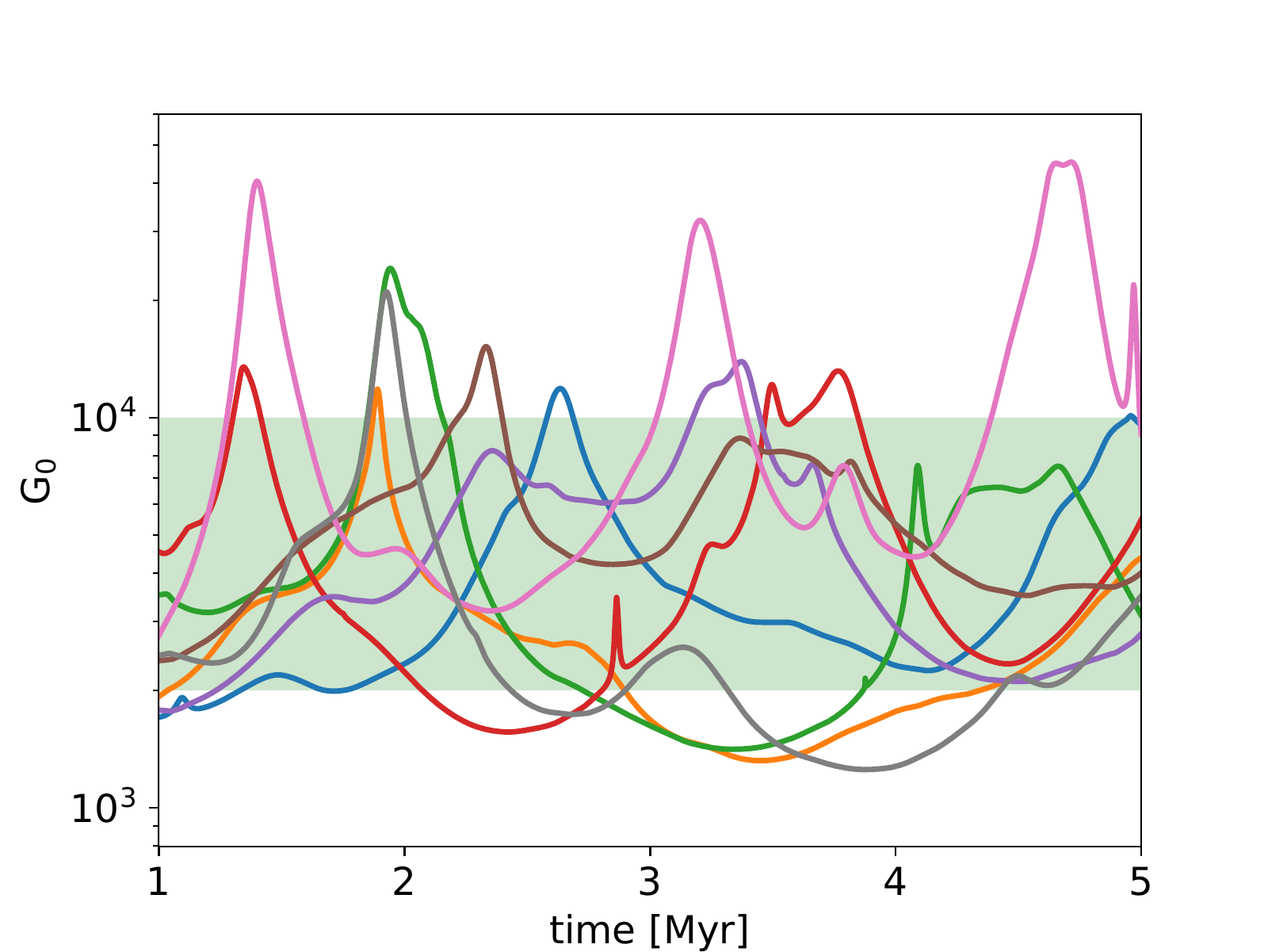}
      \caption{Irradiation tracks of eight randomly sampled G-type stars (identified by stellar mass 0.8-1.04 $M_{\odot}$) where the median background radiation field satisfies the criterion of \mbox{$2000 \leq G_0 \leq 10^4$}. The green bar indicates the boundaries of this constraint. Incident FUV radiation field strength is measured on the ordinate in units of $G_0$.  The time resolution is 1 kyr.}
      \label{fig:g0hist}
\end{figure}

The 2500 stars within the cluster are found to be exposed to a range of intensities of the intracluster radiation field from \mbox{$G_0$ = $10^0-10^7$}, with a modal \mbox{$G_0 = 3150^{+12500}_{-2400}$}.   In Fig. \ref{fig:t_g0} the temporal evolution of the distribution of intracluster FUV fluxes incident on each cluster star over the 10 Myr of the simulation is demonstrated.  We find a distribution that largely agrees with the distribution suggested by \citet{2010ARA&A..48...47A}, with a typical $G_0 \sim 2800$ and a few systems experiencing prolonged intervals of $G_0 > 10^4$. In a snapshot at $t = 0$ Myr, only $12 \%$ of our cluster stars experience $G_0 > 10^4$, while at $t = 10$ Myr, this number is reduced to $8 \%$.

Over 10 Myr, dynamical friction segregates the cluster star populations radially by mass and causes it to expand.   As lower mass stars are ejected from the cluster center where repeated close ($<0.1$ pc) encounters with the high mass \mbox{($M > 25 M_{\odot}$)} stars of the cluster occur, we find that the median $G_0$ field strength experienced by the cluster stars is reduced from an initial 2300 at $t = 0$ Myr (the red line in  Fig. \ref{fig:t_g0}) to 1100 at \mbox{$t = 10$ Myr} (the blue line).  The model $G_0$ declines from an initial 2600 to 1030. The radial intensity of the radiation field at some cluster radius $r_{\rm c}$ can be described initially by $G_0(r_{\rm c}) = 0.07 r_{\rm c}^{-2.04}$ with $r_{\rm c}$ in parsec for $r_{\rm c} > 0.05$ pc. The fraction of time that a given star spends irradiated by an FUV field $F(G_0)$ over the simulation time is found to be largely independent of $G_0$ for values of $10^0-10^2$  because these stars are found in the outer reaches of the cluster and their radial positions evolve slowly. For $G_0 > 10^2$ , however, we find that $F(G_0) \propto G_0^{-2}$. 

We traced the FUV field strength incident on G-type stars (defined as those with mass 0.84 < $M_*$ < 1.15 $M_{\odot}$) over 10 Myr to investigate the influence of stellar mass on the irradiation history of a system within the cluster. \citet{2010ARA&A..48...47A} described a constraining range of $G_0$ values necessary to explain the compact architecture of the Solar System while ensuring that the solar nebula can survive over 3-10 Myr \citep{2010ARA&A..48...47A}. A representative sample of these tracks is shown in Fig. \ref{fig:g0hist}. While only $11\%$ of the G-type stars strictly satisfy the criterion, in practice, brief excursions outside the $G_0$ range would be consistent with the physical-chemical structure of the Solar System. For a looser constraint that allows very brief ($10^4$ yr) periods of $G_0 > 10^4$, $17\%$ of the G-type stars satisfy the constraint, while $28\%$ of the median
background radiation field of the G-type star falls within the constraint. Periods of heavy irradiation ($G_0 > 10^4$) characteristically last 200-300 kyr as the stars pass rapidly through the inner regions of the cluster.  Incidents with higher irradiation ($G_0 > 10^5$, not depicted in Fig. \ref{fig:g0hist}) occur even more briefly on timescales 50-100 kyr at most.   Only $\sim20\%$ of the G-type stars ever experience radiation fields in excess of $10^6$, for which the characteristic duration is $\sim10-20$ kyr.  Half of the G-type stars that undergo the $G_0 > 10^6$ irradiation events are ejected from the cluster before 10 Myr in three-body interactions with massive stars.

\subsection{Conditions within the Jovian gap}

\begin{figure}
  \includegraphics[width=\textwidth/2]{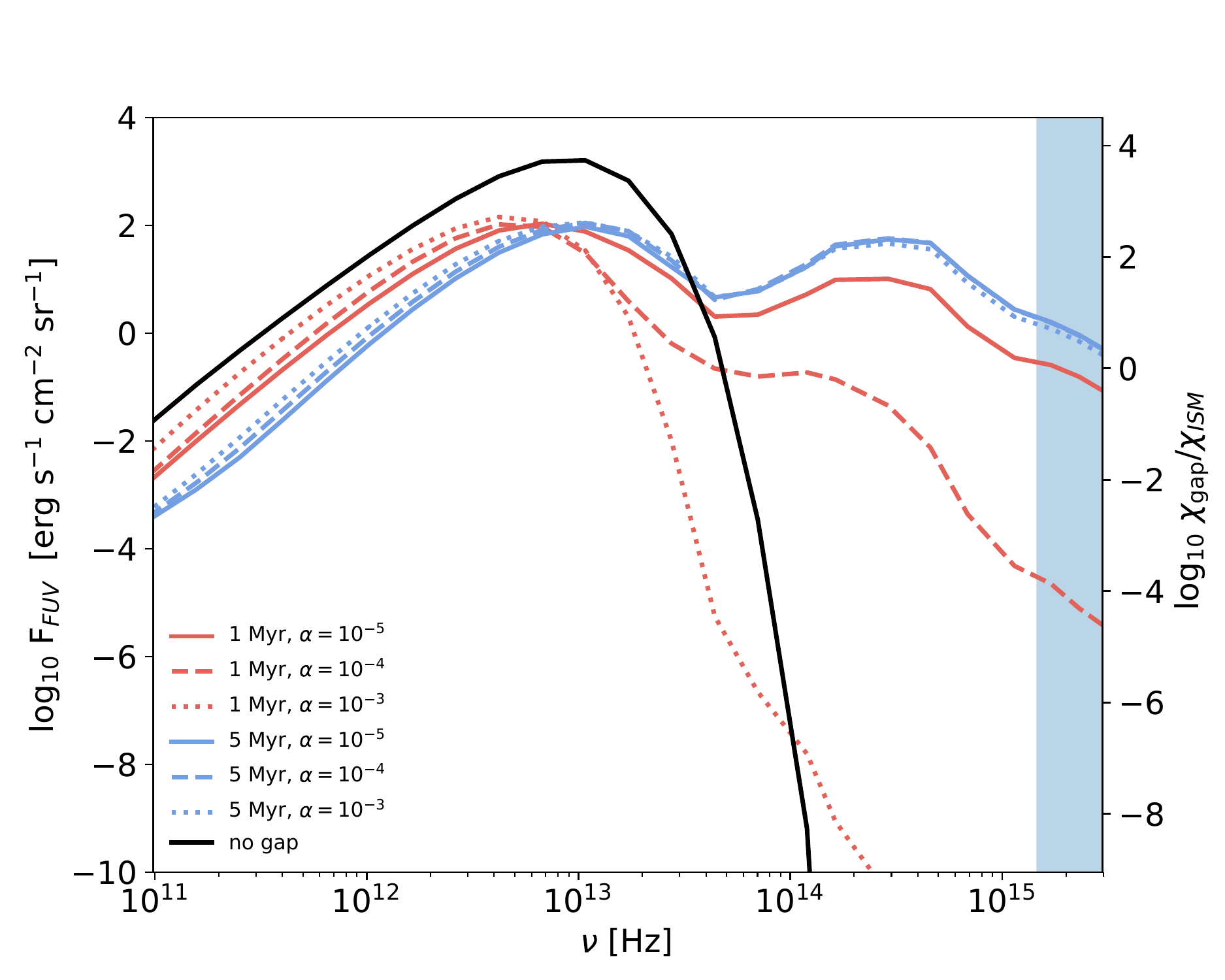}
      \caption{Extracted energy distributions of the  radiation field in the MMSN midplane at the location of Jupiter for the six gap profiles illustrated in Fig. \ref{fig:sdps}, for the case of $G_0$ = 3000.    The shaded light blue region demarcates the FUV region of the spectrum. For comparison, the black line represents the radiation field of the unperturbed solar nebula at the midplane, and hence at the surface of the CPD.  The right ordinate indicates the ratio between the external FUV radiation field originating from the cluster and that found within the gap.}
      \label{fig:radfield}
\end{figure}

The radiation field intensity within the gap for each surface density profile was extracted and is shown in Fig. \ref{fig:radfield} for a reference $G_0$ = 3000. The ratio between the FUV radiation strength within the gap and the intracluster FUV radiation can be read on the right ordinate.  The gap remains optically thick for cases with $t = 1$ Myr and $\alpha = 10^{-3}, 10^{-4}$ .  In the scenario with $t = 1$ Myr and $\alpha = 10^{-5}$ , the gap is marginally optically thin, but the interstellar FUV is still extincted by a factor 6.  The gap is highly optically thin for all $t =  5$ Myr models, allowing $>99\%$ of the vertically incident intracluster FUV to penetrate to the PPD midplane, and hence to the surface of the CPD. 

We also analyzed the radiation field inside the gap in the absence of the intracluster radiation field where the gap is illuminated only by the young Sun.  We compared the solar radiation penetrating to the PPD midplane in the six surface density profiles of Fig.\ref{fig:sdps}.  After 5 Myr, when Jupiter has grown to its final mass and the gap profiles reach their maximum depth, we find that the $G_0$ field strength within the three gap profiles is $(L_{\rm UV,*}/0.01L_{\odot}) \times 10^{3.51}, 10^{3.73}, 10^{3.84}$ respectively.  Hence, when $L_{\rm UV} = 0.01 L_{\odot}$ and $a_{\rm p}$ = 5.2 au, we find that the young Sun still contributes significantly to the intragap radiation field even in the presence of an optically thick inner disk due to the scattering of photons by the upper layers of the inner disk.  For this adopted $L_{\rm UV,*}$  the contribution of the young Sun to the FUV background is greater than that of the intracluster radiation field for $82\%$ of stars in the cluster at 5 Myr.  The cluster thus only becomes the dominant source of FUV radiation for stars with this disk-gap configuration if $L_{\rm UV,*} < 2.6 \times 10^{-3} L_{\odot}$.  

For a nominal $G_0 = 3000$ at $t = 1$ Myr, we find midplane gas temperatures within the gap of 50-65 K for the $\alpha = 10^{-3}, 10^{-4}$ , and $10^{-5}$ cases.  In the $t$ = 5 Myr scenarios where the Jovian gap becomes optically thin and volume number densities approach $10^{7}$ cm$^{-3}$, the intragap gas temperature ranges between \mbox{$96 - 320$ K} in the midplane due to X-ray Coulomb and photoelectric heating for $G_0 = 3000$.  For the maximum $G_0 = 10^6$ ,we find a midplane gas temperature of 5000 K driven by polycyclic aromatic hydrocarbon (PAH) heating. In all cases, efficient \ion{O}{i} line cooling leads to a torus of cooler gas of $T < 100$ K suspended above and below the midplane.   The cool torus extends vertically and reaches the midplane for the \mbox{$t = 5$ Myr}, $\alpha = 10^{-5}$ case.

The primary source of gap-wall radiation that the CPD is exposed to is thermal emission from the dust. The dust temperature $T_{\rm dust}$ of the gap walls near the midplane at a radial optical depth of 1 is calculated by  \textsc{ProDiMo} 2D radiative transfer. This temperature can be considered as the blackbody temperature of the gap walls, and thus the "background temperature" of the CPD.  In all cases where the gap is optically thin, we find gap wall temperatures ranging from 60-75 K. The gas and dust temperature structure of the solar PPD for the case of $G_0= 10^3$, $t = 5 $ Myr, and $\alpha = 10^4$   is described in Fig. \ref{fig:mmsn_all} in the appendix.The outer gap wall is consistently 10 K warmer than the inner gap wall.  We find that the optically thick surfaces of the gap walls are largely insensitive to increasing $G_0$ for $G_0$ up to $10^{5}$, while for $G_0 = 10^{6}$ , we find a general increase of 20 K for the inner and outer gap walls.

\subsection{Temperature structure and truncation of the CPD}

We generated \textsc{ProDiMo} CPD models for intracluster FUV intensities spaced logarithmically in the range $G_0$ = $10^1-10^7$.  We considered five CPD models of different masses (see Table \ref{tab:cpds}).  We exposed them to a range of $G_0$ spanning seven orders of magnitude. We determined mass-loss rates based on the gas sound speed of the 35 resulting models. We interpolated over our model grid to derive mass-loss rates for the CPDs exposed to arbitrary values of $G_0$ in the range $G_0$ = $10^1-10^7$. For all modeled cluster ISRF cases we find CPD surface layers heated to $T > 10^3$ K and associated regions at which the local sound speed exceeds the escape velocity.  The gas and dust temperature structure of the fiducial CPD are shown in Fig. \ref{fig:cpd_all} in the appendix. In the highly optically thin $M_{\rm CPD} = 10^{-8}-10^{-9} M_{\odot}$ cases, we find gas loss that directly occurs from the CPD midplane.  We considered a range of steady-state mass accretion rate \mbox{$\dot{M} = 10^{-12}-10^{-9} M_{\odot}$ yr$^{-1}$}, and also the case of an exponential decline and cutoff in the accretion rate.  

Circumplanetary
disks evolve to a truncated steady state on a $10^3$ yr timescale when mass loss through photoevaporation is balanced with the mass accretion rate.  As a consequence of the orbital motion of the stars through the cluster, the CPDs in general are maximally truncated on timescales $< 10^5$ yr. For a given mass accretion rate, the instantaneous distribution of CPD truncation radii is then determined by the radial distribution of stars from the cluster center where FUV field strengths are highest. 

\begin{figure}
  \includegraphics[width=\textwidth/2]{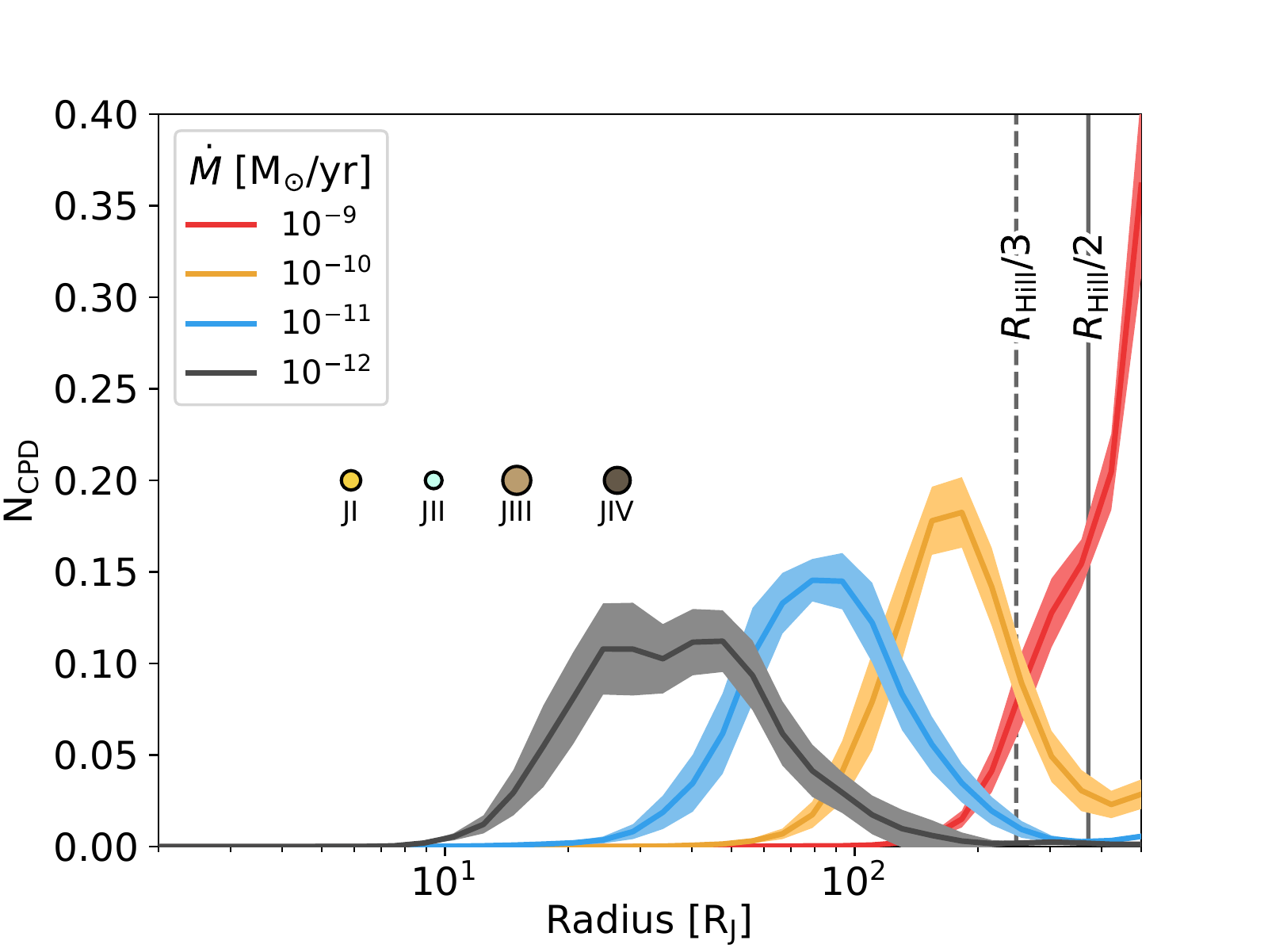}
      \caption{Distribution of steady-state truncation radii for the grid of CPD models in the case of intracluster FUV irradiation at \mbox{5 Myr}.  Each colored line represents the instantaneous distribution of the outer radius of 2500 Jovian CPD analogs with a range of external FUV radiation field strengths.  The shaded region bracketing each distribution indicates the standard deviation between the different CPD mass models.  The four colored circles indicate the radial location of the Galilean satellites JI (Io), JII (Europa), JIII (Ganymede), and JIV (Callisto). The vertical dashed black lines indicate theoretical limits on the CPD outer radius based on gravitational perturbations as fractions of the Hill radius.}
      \label{fig:truncrad2}
\end{figure}

We find only a weak dependence of CPD mass on steady-state truncation radius, while the accretion rate is found to dominate the resulting steady-state CPD radii.  The results are plotted in Fig. \ref{fig:truncrad2}.   For each sampled accretion rate, the scatter induced by the CPD mass is bracketed by the shaded regions around the solid line.  We find that for low accretion rates ($10^{-12} M_{\odot}$ yr$^{-1}$), $\sim50\%$ of the Jovian CPD analogs in our cluster are truncated to radii within 30 $R_{\rm J}$ with a modal \mbox{$r_{\rm trunc}$ = 27 $R_{\rm J}$}.   For intermediate accretion rates ($10^{-10} M_{\odot}$ yr$^{-1}$), $50\%$ of the CPDs are truncated to within 110 $R_{\rm J}$ with a modal \mbox{$r_{\rm trunc} = 200$ $R_{\rm J}$}. We find that the truncation radius is proportional to the accretion rate $\dot{M}^{0.4}$. The distribution of the truncation radius for stars that conform to the solar system formation constraint of $2000 < G_0 < 10^4$ is $28.7^{+5.4}_{-2.6} R_{\rm J}$ at $t = 5$ Myr for $\dot M = 10^{-12} M_{\odot}$ yr$^{-1}$.

The width of the truncation radius distributions in Fig. \ref{fig:truncrad2} is largely induced by the distribution of $G_0$ within the cluster, and we find that their relation is fit by

\begin{equation} \label{eq:trunc}
    r_{\rm trunc} \approx \textrm{min} \{  \: 2\times10^7 \bigg( \frac{\dot M^{0.4}}{\textrm{log}_{10}(\textrm{G}_0)^2} \bigg) \: , r_{\rm out}  \: \} \: R_{\rm J}\end{equation}

\noindent
for accretion rates $10^{-12} \leq \dot M \leq 10^{-9} M_{\odot}$ yr$^{-1}$ and FUV fields $10^1 \leq G_0 \leq 10^7$.  We find the relation between the remaining mass of the truncated CPD $M_{\rm CPD,trunc}$ to be a fraction of its initial steady-state mass $M_{\rm CPD,init}$ , and the accretion rate $\dot M$ is 

\begin{equation}
    \bigg(\frac{M_{\rm CPD,trunc}}{M_{\rm CPD,init}}\bigg) \approx \frac{1 - M_{\rm CPD,min}}{1+\exp{[-1.87 (\textrm{log}_{10}(\dot M) +10.42)]}} + M_{\rm CPD,min} \: ,
\end{equation}
\noindent
where $M_{\rm CPD,min}$ is the fraction of CPD mass that remains after accretion drops significantly below $10^{-12}$ M$_{\odot}$ yr$^{-1}$, which in the case of Jupiter, we find to be $8\%$.  For the modal FUV radiation field strength in the cluster, $15^{+6}_{-6}\%$, $34^{+14}_{-12}\%$, $69^{+15}_{-18}\%$, and $99^{+1}_{-8}\%$ of the initial steady-state mass of the CPDs thus remains for mass accretion rates  of $\dot M = 10^{-12}$, $10^{-11}$, $10^{-10}$, and  \mbox{$10^{-9}  M_{\odot}$ yr$^{-1}$}, respectively.

\begin{figure}
      \includegraphics[width=\textwidth/2]{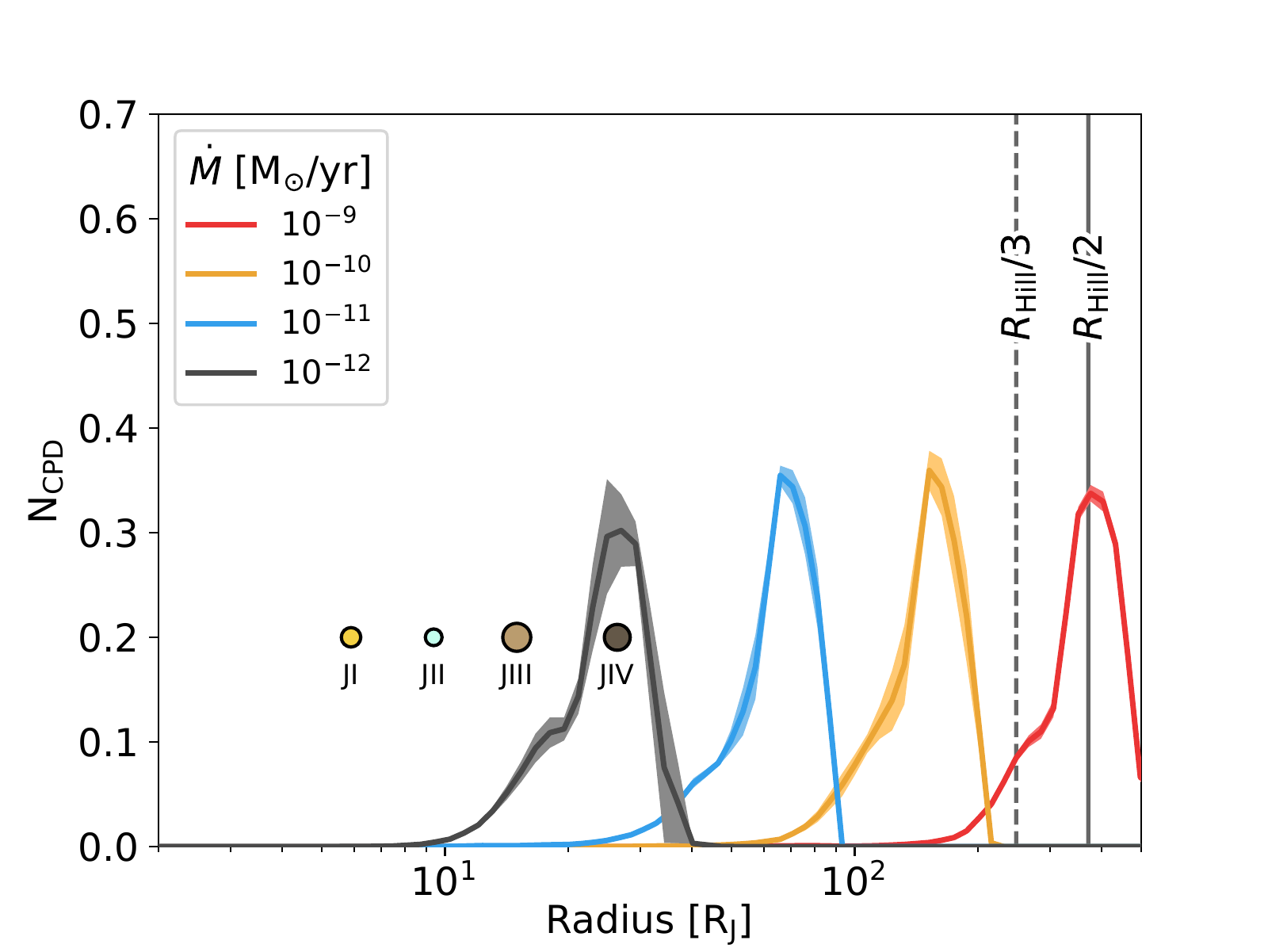}
      \caption{Same as in Fig. \ref{fig:truncrad2}, but now also including the effect of the solar radiation.}
      \label{fig:truncrad2_withSun}
\end{figure}

\subsection{Photoevaporation rates with alternative CPD and planet parameters}
 
\begin{figure}
  \includegraphics[width=\textwidth/2]{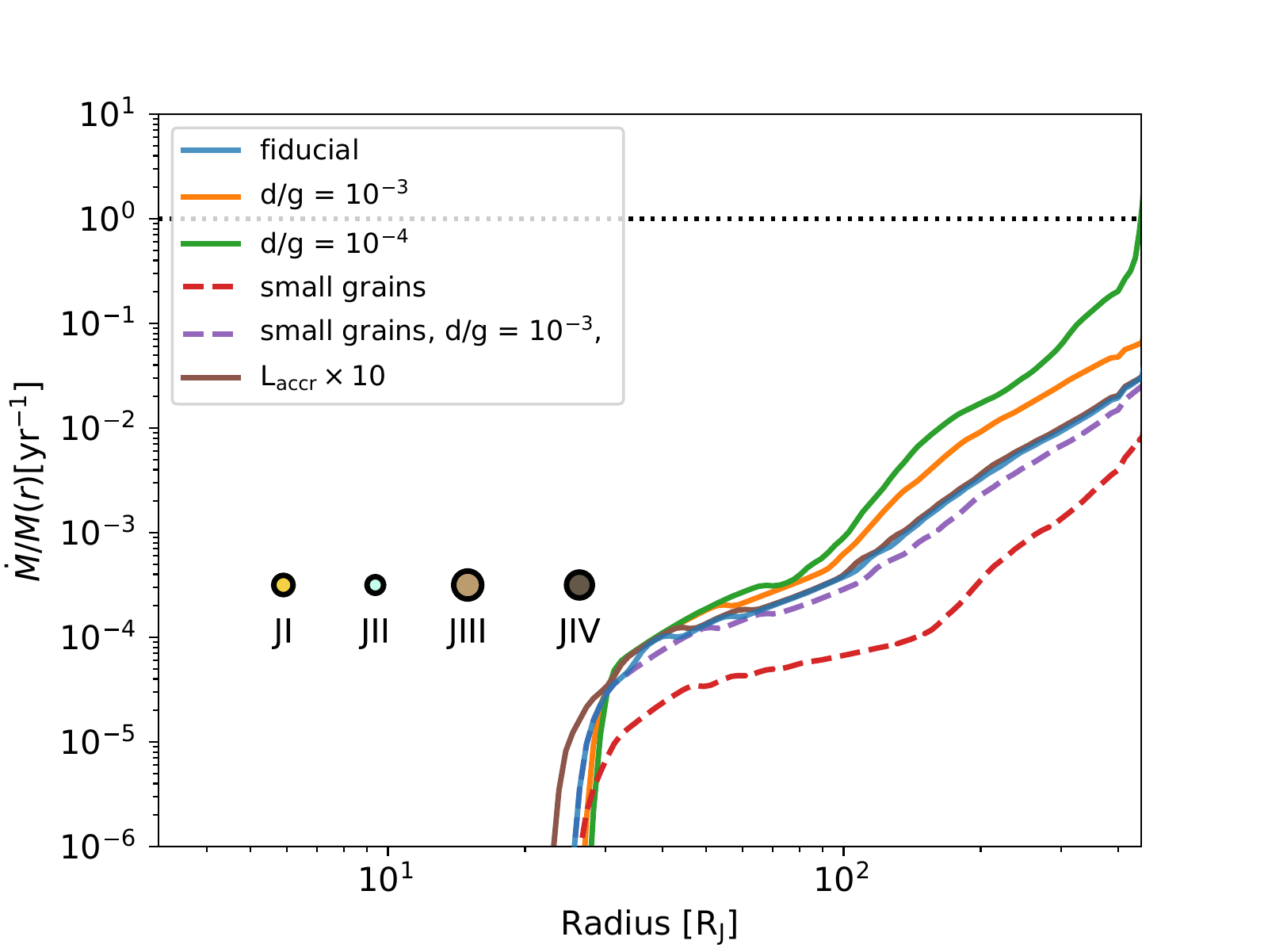}
     \caption{Radial photoevaporative mass loss in the CPD as a fraction of the mass available in the respective annulus of the steady-state  surface density profile for a variety of modifications to the fiducial CPD model ($M_{\rm CPD} = 10^{-7} M_{\odot}, G_0 = 10^3$).  The red and purple curves labeled "small grains" represent the case where the maximum grain size is 10 $\mu$m.  The orange, green, and purple curves represent the cases of varying dust-to-gas ratio.  The brown curve shows the case of additional planetary UV luminosity originating from accretion, $\text{ten times}$ greater than in the fiducial case of planetary UV luminosity. The horizontal dashed black line represents the mass-loss rate at which an annulus would be entirely depleted within one year.  The colored circles indicate the current semimajor axes of the Galilean satellites as described in Fig. \ref{fig:truncrad2}.}
      \label{fig:mlossperdr}
\end{figure}

Pressure bumps can act to filter and segregate dust-grain populations based on grain size \citep{2006MNRAS.373.1619R,2012ApJ...755....6Z}.  We also considered the case of a modified grain size distribution in the CPDs and how it changes the rate of mass loss to photoevaporation.  For the grain-filtered scenario, the maximum dust grain size was set to 10 $\mu$m rather than 3 mm \citep{2007A&A...462..355P, 2012ApJ...755....6Z, 2018A&A...612A..30B}. Because the material supplying the CPD may thus be starved of large dust grains, we also varied the dust-to-gas ratio.  We considered disks of $d/g = 10^{-2},10^{-3},\text{and }10^{-4}$.  The resulting radial mass-loss profiles are shown in Fig. \ref{fig:mlossperdr}.  The fiducial CPD model we present is the $M_{\rm CPD} = 10^{-7} M_{\odot}$, with a background radiation field $G_0 = 10^3$ and $d/g = 10^{-2}$.  A lower dust-to-gas ratio decreases the disk opacity and pushes the FUV-heated envelope closer to the midplane, where volume density and thus mass-loss rates are higher.  The radial mass loss  $\dot M(r)$ is related to the dust-to-gas ratio $d/g$ as $\dot M(r) \propto (d/g)^{0.56}$.  The removal of larger (> 10 $\mu$m) grains conversely suppresses the photoevaporation as the mass previously stored in the large grains is moved to the small grains where the bulk of the opacity lies.  In the combined case of both small grains and a reduced dust-to-gas ratio, the radial mass-loss rate of the fiducial case is closely reproduced, as is shown by the agreement of the blue and purple curve in Fig. \ref{fig:mlossperdr}.  

We also considered cases of increased planetary UV luminosity driven by the higher rates of mass accretion. For the case of $L_{\rm UV,p} = 10^{-4} L_{\rm p}$  , we find that the highly optically thick CPDs of mass $M_{\rm CPD} \geq 10^{-6} M_{\odot}$ are insensitive to the increased planetary luminosity.  The lower mass CPDs experience both an increase in mass loss at radii within 200 $R_{\rm J}$ and mass loss in the innermost radii (see the brown curve in Fig. \ref{fig:mlossperdr}).  The minimum of the truncation radius distribution is then pushed to lower radii.  While an arbitrarily massive CPD exposed to $G_0 > 10^3$ is not truncated inward of $20$ $R_{\rm J}$, increasing the planetary luminosity up to \mbox{$10^{-3}$ L$_{\odot}$} can decrease the minimum truncation radius to within 6 $R_{\rm J}$, within the centrifugal radius at which mass is expected to accrete.  We do not expect planetary luminosities this high at the late stage of satellite formation, however.

\subsection{Photoevaporative clearing}

We considered how rapidly photoevaporation can act to clear the CPD. Disk-clearing timescales are of particular significance for the fate of satellites undergoing gas-driven migration, with implications for the final system architecture.   Jovian planets that are starved of accretion material, perhaps by the formation of adjacent planets, may lose most of their mass through rapid photoevaporative clearing \citep{2011AJ....142..168M}. A rapid cutoff of the accretion may occur when the gaps of Saturn and Jupiter merge to form a single deeper gap, abruptly starving Jupiter of material originating external to its own orbit \citep{2010ApJ...714.1052S}.  We considered the CPD lifetimes in the context of such a rapid accretion cutoff.

The maximum photoevaporative mass-loss rate occurs for the $10^{-5} M_{\odot}$ CPD when exposed to the maximum considered radiation field $G_0$ = $10^7$. For a static nonaccreting $10^{-5} M_{\odot}$ CPD, we then find a minimum lifetime against photoevaporation of $5\times10^4$ yr.  In practice, the CPDs are rarely exposed to radiation fields greater than $G_0$ = $10^5$ for extended periods of time, with a maximum exposure time of $\sim 10^5$ yr (see Sect. \ref{sec:isrf}).  For the most likely value of $G_0$ $\approx 10^3$ , we find a static CPD lifetime of $\tau_{\rm disk}$ = $4\times10^5$ yr against intracluster photoevaporation.  In the case of the low-mass optically thin $10^{-9} M_{\odot}$ CPD, we find a disk lifetime against photoevaporation of only 25-300 years. The upper boundary corresponds here to the maximum possible initial CPD outer radius and hence lowest background radiation field strength observed in the cluster of $G_0 \sim 10^1$.

\begin{figure}
  \includegraphics[width=0.5\textwidth]{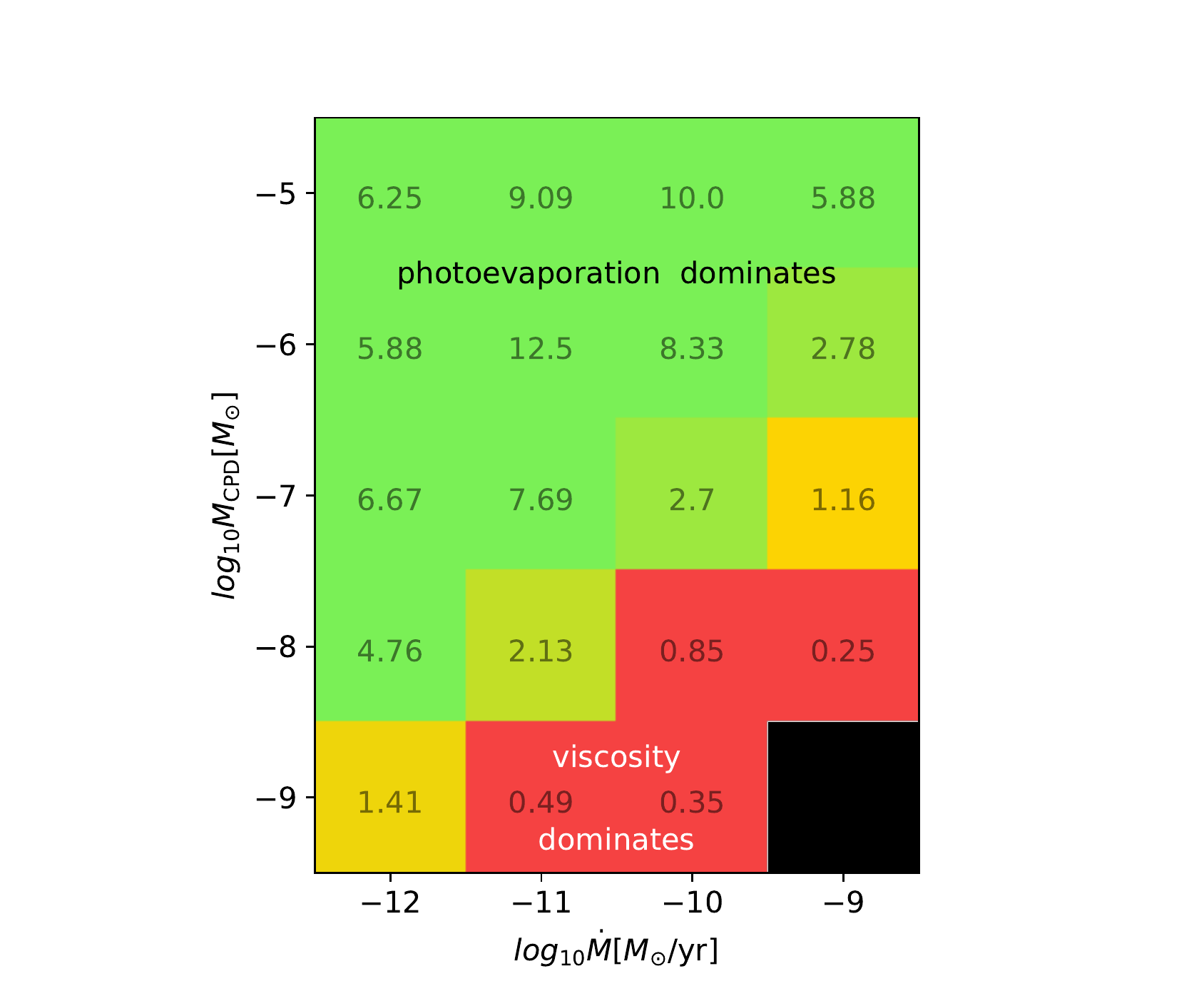}
      \caption{Ratio of the viscous diffusion timescale $\tau_{\rm visc}$ over the photoevaporative clearing timescales $\tau_{\rm evap}$ for the grid of CPD mass and accretion rate models.   A value greater than 1 indicates that photoevaporation clears the disk more rapidly than viscous evolution. The black region indicates an unphysical corner of the parameter space.}
      \label{fig:tabletime}
\end{figure}

In the absence of photoevaporation, the CPD will dissipate on its viscous diffusion timescale. In Fig. \ref{fig:tabletime}, we show which regions of the CPD mass and accretion rate parameter space allow for photoevaporation to be the dominant disk-clearing mechanism.  Viscous clearing primarily dominates for the high-mass ($\geq 10^{-8} M_{\odot}$) CPD models with higher mass accretion rates ($\geq10^{-10.5} M_{\odot}$ yr$^{-1}$).  Realistically, viscous evolution of the CPD will act in tandem with photoevaporation, transporting material to the more weakly bound regions of the CPD where photoevaporation is far more efficient \citep{2011AJ....142..168M}.

\section{Discussion}

\subsection{Relevance of photoevaporation for CPD size and lifetime}

We have found that the size of a CPD is directly proportional to the mass accretion rate onto the CPD and FUV radiation field strength by Equation \ref{eq:trunc}.  The photoevaporation mass-loss rate of the CPDs has been found to be sufficiently rapid to enable a clearing of the CPD in only $ < 10^3$ yr for the low-mass ($< 10^{-7} M_{\odot}$) CPD cases. A truncated outer radius and a rapid clearing of the CPD at the end of satellite formation has direct implications for the architecture of late-forming satellite systems.   Unless the accretion onto the CPD is abruptly reduced to $\dot M \ll 10^{-12} M_{\odot}$ yr$^{-1}$ from a previous steady accretion rate $\dot M \geq 10^{-11}$, the Jovian CPD will most likely be truncated to \mbox{30 $R_{\rm J}$} at some stage during or prior to satellite formation.  

We find that solar (intrasystem) FUV radiation can still contribute significantly to CPD photoevaporation compared to the cluster FUV field.  In the case of the optically thin gap allowing for full exposure of the CPD, we find that solar radiation places a lower limit on disk truncation regardless of the position of the system within the cluster, as illustrated in Fig. \ref{fig:truncrad2_withSun}.  For the purpose of this figure, each star was assumed to have solar FUV luminosity, therefore it does not represent the distribution of truncation radii arising from the luminosity function of stars within the cluster, but rather the possible positions of the Sun within the cluster.

However, we recommend a more sophisticated radiative transfer model that includes the effects of anisotropic scattering of light by dust grains to determine the effect of the CPD inclination on the efficiency of this solar irradiation.  This method would also be able to more accurately probe the radiation contribution of the gap walls, which are heated by the Jovian luminosity.

\subsection{Comparison with previous work}

In their study of the Jovian CPD, \citet{2011AJ....142..168M} constructed a 1D model that coupled viscous evolution, photoevaporation, and mass accretion for a range of fixed CPD envelope temperatures.  They found a distribution of Jovian CPD truncation radii ranging from 26-330  $R_{\rm J}$ with a mean value of $\sim120 R_{\rm J}$.   To reproduce the same mean truncation radius, our model requires a slightly lower accretion rate of $\sim10^{-10.4} M_{\odot}$yr$^{-1}$ compared to the $\dot M \approx 10^{-9.9} M_{\odot}$ yr$^{-1}$ considered by \citet{2011AJ....142..168M}, although this can likely be explained by different choices of CPD surface envelope temperatures.  Our truncation radii are found to be tied only to the background radiation field strength and the mass accretion rate, but not significantly to the CPD mass and hence not to the viscosity. This is consistent with the findings of \citet{2011AJ....142..168M} that the $\alpha$-viscosity alone does not significantly alter the truncation radius.  

\citet{2011AJ....142..168M} also studied photoevaporative CPD clearing timescales, finding evaporation times $\tau_{\rm evap}$ of $10^2-10^4$ yr, with the spread arising from the range of  envelope temperatures 100-3000 K.  For our fiducial $G_0 = 3000,$ we find a corresponding CPD surface envelope temperature of \mbox{1900 $\pm$ 100 K}. For CPDs of similar viscosity and mass accretion rate than they considered ($\alpha = 10^{-3}$, $\dot M \approx 10^{-9.9} M_{\odot}$yr$^{-1}$), we find $\tau_{\rm evap} = 3\times10^2 - 10^3$ yr, with the spread caused by the range of possible initial CPD outer radii.  Our results thus appear to be largely consistent for CPDs of similar surface temperatures. 


\subsection{Implications of photoevaporative truncation}

In the case $\dot M \leq 10^{-12} M_{\odot}$ yr$^{-1}$ , we find that photoevaporative truncation provides a natural explanation for the lack of massive satellites outside the orbit of Callisto. The truncation of the CPD would cause the satellite systems of $\sim 50\%$  Jupiter-mass planets forming in star clusters of N $\sim 2500$ to be limited in size to 0.04-0.06 $R_{\rm H}$.  The outermost extent of these Galilean analog systems would follow the truncation distribution of the low-accretion case in Fig. $\ref{fig:truncrad2}$. 

Circumplanetary
disk truncation could also act to bias our interpretation of unresolved continuum point-sources suspected to be CPDs.   In the optically thin case, fluxes are used to infer CPD mass independent of the physical dimensions of the CPD, while in the optically thick case, the inferred radius of the CPD is proportional to the flux $F_{\nu}$ and dust temperature $T_{\rm d}$ by $r_{\rm CPD}^2 \propto F_{\nu} B_{\nu}(T_{\rm d})^{-1} $ \citep{2019ApJ...871...48P}.  Our results show that for low accretion rates, CPD outer radii may be as small as 0.04 $r_{\rm H}$ rather than \mbox{0.3-0.5 $r_{\rm H}$} \citep{1998ApJ...508..707Q, 2011ESS.....2.3311M, 10.1111/j.1365-2966.2009.15002.x, 2013ApJ...767...63S}.  The dust temperature of a CPD observed at a given flux might then be overestimated by a factor 3-5, with direct implications for the inferred luminosity of the planet.

We also placed limits on the strength of the intracluster radiation field $G_0$ during the formation time of the Galilean satellites for certain rates of mass accretion.  If the Galilean satellites formed late during a period of slow accretion \mbox{($\dot M \leq 10^{-12} M_{\odot}$ yr$^{-1}$)}, we placed upper limits on the FUV field strength \mbox{$G_0 \leq 10^{3.1}$} that would still allow the satellites to form at their present-day locations due to the correspondingly small CPD truncation radius.  The semimajor axis of the Saturn moon Titan would need to be explained in this context as either forming at a later stage in an epoch of lower $G_0$ or forming much closer in to Saturn initially and migrating outward as a result of tidal dissipation.  A close-in formation scenario for Titan has been proposed based on the spreading of tidal disks \citep{2012Sci...338.1196C}, and high tidal recession has been observed in the Saturnian system, possibly necessitating a close-in formation scenario \citep{2012A&A...541A.165R, 2017Icar..281..286L,2019EPSC...13.1685G}. With our model applied to a Saturn-mass planet, we find that the inner mean truncation radius corresponding to $\dot M = 10^{-12} M_{\odot}$ yr$^{-1}$ shrinks to \mbox{$\sim 8.3$ $R_{\rm S}$}. If Titan formed within this radius, it would necessitate a migration of at least 12.7 $R_{\rm S}$ to approach its current position.  With the resonant-locking mechanism, \citet{2016MNRAS.458.3867F} suggested a tidal migration timescale defined as $a / \dot a$ for Titan of $\sim$2 Gyr.  If we were to naively assume a fixed tidal migration timescale this might suggest a change in semimajor axis of  $\delta a_{\rm Titan} \approx$ 13.5 $R_{\rm S}$ over 4.5 Gyr for a final $a_{\rm Titan} \sim$ 19.5 $R_{\rm S}$ for an initial $a_{\rm Titan} = 6$ $R_{\rm S}$ . This is near the current semimajor axis of 20.27 $R_{\rm S}$.  We thus find that the photoevaporative truncation of the Saturn CPD is consistent with a close-in formation scenario for Titan and its subsequent outward migration.

\subsection{Photoevaporative stranding of Callisto}\label{sec:chain}

Rapid clearing of the gas in CPDs could function to prevent the inward migration of satellites already present in the CPD.  The inward migration of resonant chains of planets has been suggested as a mechanism to explain compact resonant exoplanet systems such as TRAPPIST-1 \citep{2017A&A...604A...1O,2017MNRAS.470.1750I,2017ApJ...840L..19T}.  It appears to be a general outcome of planet formation models around low-mass stars that super-Earths migrate inward and form compact systems of resonant chains \citep{2007ApJ...654.1110T,2008MNRAS.384..663R,2010MNRAS.401.1691M,2015ApJ...798...62L}, although in situ formation is possible \citep{2018AJ....156..228M}.

Callisto is noteworthy in that it does not lie within the Laplace resonance of the inner three Galileans. Previous mechanisms proposed to explain why Callisto is excluded from the resonance have included its insufficiently rapid inward orbital migration  \citep{Peale593,2002AJ....124.3404C},  perturbation out of an initial resonance by dynamical encounters between Jupiter and other planets \citep{2014AJ....148...25D}, or scattering by close encounters with Ganymede \citep{2019ApJ...885...79S}.  While the satellites may also have evolved away from their initial positions by secular processes into their current positions \citep{1979Natur.279..767Y, 2016MNRAS.458.3867F, 2019EPSC...13.2017L}, a primordial formation of the current configuration by differential orbital migration and tidal dissipation is also possible \citep{2002Sci...298..593P}.  

The observed paucity of intact compact resonant chains in exoplanetary systems suggests that these systems undergo an instability to break the resonant configurations \citep{2014AJ....147...32G, 2017MNRAS.470.1750I}.  The instabilities may be caused by the dissipation of the gas disk in which the planets migrated  \citep{2009ApJ...699..824O,2014A&A...569A..56C}.  The Galilean Laplace resonance would need to have remained stable during the CPD gas-disk dispersal phase. Resonant chains with the lowest number of planets in the chain and lowest planet masses are the most stable \citep{2019arXiv190208772I,2012Icar..221..624M}. 

When we assume that the Galilean satellites migrated primordially into the 4:2:1 mean motion resonance by gas-driven inward migration and retained the Laplace resonance during dispersal of the gas, the question arises whether a rapid photoevaporative truncation of the CPD might have stranded Callisto by preventing further migration to complete the chain, even for a high-mass CPD.  We first investigated the relative timescales of the photoevaporative disk clearing and of the orbital migration.  We considered the case of the Laplace resonance being primordial (coeval with the formation of the satellites themselves), and that the resonance survived the dispersal of the gas disk.  Additionally, we focused in detail on the case of an abruptly terminated mass inflow rather than a slow tailing-off of the mass accretion rate. 

\begin{figure*}
  \includegraphics[width=\textwidth]{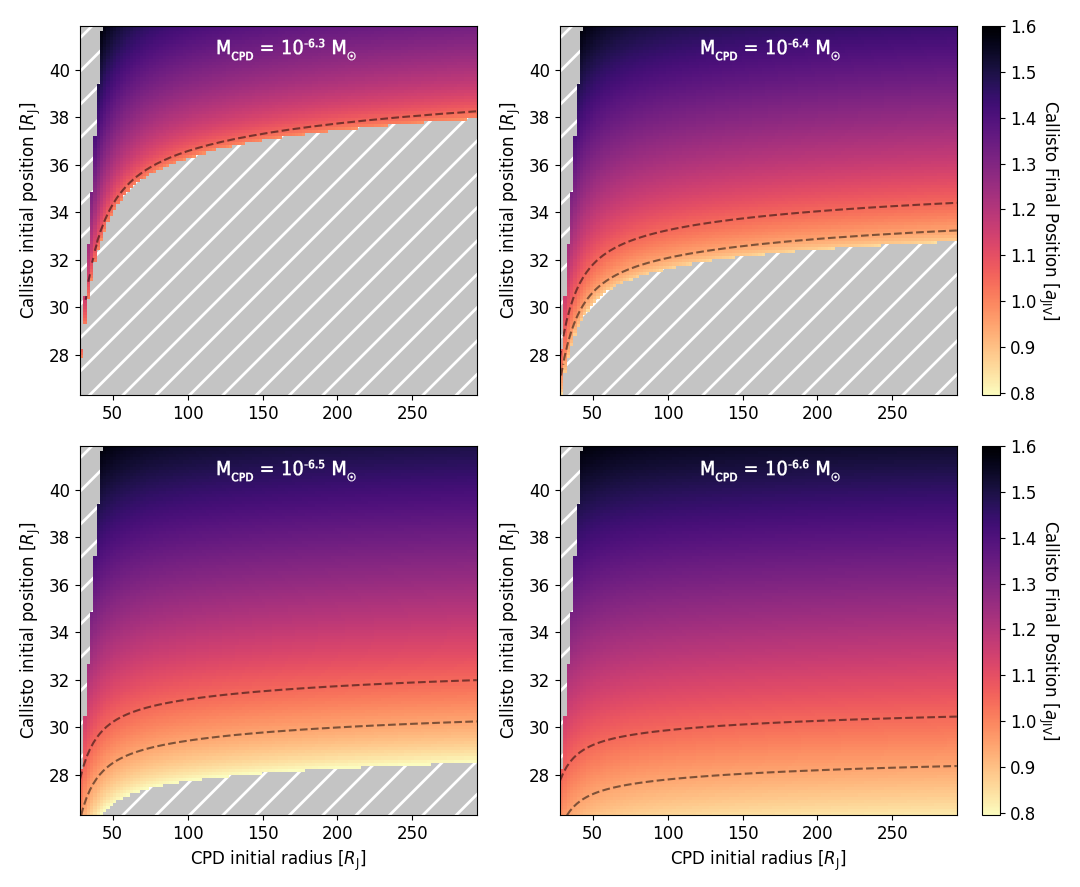}
      \caption{Final semimajor axis of Callisto $a_{\rm final}$ in our migration and photoevaporation model for the case of $M_{\rm CPD} = 10^{-6.3}$ (top left), $10^{-6.4}$ (top right), $10^{-6.5}$ (bottom left), and $10^{-6.6} M_{\odot}$ (bottom right).  The initial radius of the CPD is on the ordinate, and initial location of Callisto is on the abscissa. The region of excluded parameter space where Callisto already lies beyond the outer edge of the CPD and where Callisto enters the resonance is filled with gray stripes.   The dotted black lines demarcate the interval where the final position of Callisto falls within $5\%$ of its present-day position, $a_{\rm final} = a_{\rm JIV}$.}
      \label{fig:captureplot}
\end{figure*}

Type I migration is most efficient when the gas disk density is high at small orbital radii and short orbital periods. In planet formation models,  efficient type I migration causes `fast' migration, which can result in scattering and collisions that leave between four to six planets in mean motion resonances both before and after the gas disk dissipates \citep{2009ApJ...699..824O}.  Superficially, this scenario corresponds to the final configuration of the Galilean satellites.  Slower migration results in very efficient resonant capture and the formation of long chains that are broken only when the gas disk dissipates and the effects of eccentricity damping from tidal interaction no longer plays a role \citep{2009ApJ...699..824O}.  We considered that owing to the compact nature of the CPD and the very high ($> 10^5 $ g cm$^{-2}$) surface densities, type I migration may be highly efficient in the Jovian CPD, such that the fast migration scenario is more likely.  If MRI-induced turbulence in the disk is limited to the CPD upper layers, the protosatellites may have orbited within a dead-zone dominated by laminar flow. In this case, the type I migration timescale of a satellite is described by the ratio of the orbital angular momentum $J_{\rm s}$ and the total disk-exerted torque $\Gamma_{\rm tot}$ of the object.  To semianalytically determine the satellite migration timescales, we adopted the relations of \citet{2012ARA&A..50..211K}, where the migration timescale $\tau_{\rm mig}$ is

\begin{equation}
\tau_{\rm mig} = \frac{1}{2} \frac{J_{\rm s}}{\Gamma_{\rm tot}}
,\end{equation}

\noindent
with the orbital angular momentum defined as
\noindent
\begin{equation}
J_{\rm s} = m_{\rm s} \sqrt{GM_*a_{\rm s}} \: .
\end{equation}

\noindent
Here $m_{\rm s}$ and $a_{\rm s}$ are the satellite mass and semimajor axis.  The total torque is then
\noindent
\begin{equation}
\Gamma_{\rm tot} = -(1.36 + 0.6 \beta_{\Sigma} + 0.43\beta_T) \Gamma_0 \: ,
\end{equation}

\noindent
where $\beta_T$ and $\beta_{\Sigma}$ are the power-law exponents of the temperature and surface density profiles, respectively, and the torque normalization is
\begin{equation}
    \Gamma_0 = \frac{m_s^2}{M_p} \frac{H}{r_s}^{-2} \Sigma_{\rm s} r_s^4 \Omega_{\rm s}^2 \: .
\end{equation}

Here $M_{\rm p}$ is the planet mass, $r_{\rm s}$ and $\Omega_{\rm s}$ are the satellite orbital radius and orbital angular frequency, $H/r$ is the ratio between the disk scale height at some radius $r$ and $r$, and $\Sigma_{\rm s}$ is the surface density at the orbital radius of the satellite.  We extracted the calculated temperature profile and the parameterized surface density and scale height profiles from our \textsc{ProDiMo} models to calculate migration timescales for all radii and arbitrary satellite masses.  We then compared the possible migration paths of Callisto with the disk-clearing timescales determined from our photoevaporation model.  

For the initial conditions, we sampled from a range of semimajor axes for Callisto and a CPD outer edge from 26-500 $R_{\rm J}$.  We then calculated the inward motion of Callisto through the \mbox{type I} migration rate with the  $\Sigma$ and $\beta_{\rm T}$ parameters extracted from our \textsc{ProDiMo} photoevaporation CPD model output. $\beta_{\Sigma}$ and $H(r)$ are parameterized inputs. We neglected gravitational interaction between satellites, nongravitational effects such as aerodynamic drag, and the possibility of type II migration.  The rate of photoevaporative shrinking of the CPD was calculated from our truncation model, where we exponentially decayed the mass accretion rate on a short $10^3$ yr timescale and allowed photoevaporation to evolve the CPD away from its steady-state surface density profile.  We traced both the position of Callisto and the CPD outer radius simultaneously as a function of time.  We determined whether the outer edge of the CPD moves inward past Callisto as it migrated (in which case it is stranded), or whether it would be able to migrate into the resonant chain.  

It is evident that for a migrating Callisto to be stranded by a rapidly dispersing disk, its migration timescale $\tau_{\rm mig}$ at some radius $r$ must be longer than the disk-clearing timescale $\tau_{\rm evap}$ at that radius.  While the type I migration accelerates as Callisto moves into the denser regions of the CPD and increases its orbital frequency,  photoevaporation slows as the gas at the shrinking outer radius becomes more gravitationally bound.  This means that for each CPD model, a critical radius exists inside of which Callisto can no longer be stranded by photoevaporation. We first considered the mass of proto-Callisto as a free parameter.  A CPD of mass $10^{-5} M_{\odot}$ results in a type I migration timescale so rapid that even a low-mass proto-Callisto ($M_{\rm JIV}$ = 0.01 $M_{\rm final}$) cannot be caught by disk photoevaporation at its present-day location.  In the case of the $10^{-6} M_{\odot}$  CPD, we find that the proto-Callisto could only have accreted $<10\%$ of its final mass in order to be caught by photoevaporative disk dispersal, but is highly unlikely to be placed at \mbox{26 $R_{\rm J}$} for any considered initial conditions.  This scenario would require Callisto to accrete $>90\%$ of its mass after the dispersal of the gas-disk. We consider this scenario not plausible. 

When we instead consider a scenario in which Callisto has acquired a minimum of $90\%$ of its final mass prior to the CPD dissipating, we can constrain the CPD masses that allow for Callisto to be deposed at its current orbital location. The final position of Callisto as a function of its initial position and CPD outer radius for the cases $M_{\rm CPD} = 10^{-6.3}-10^{-6.6} M_{\odot}$ is plotted in Fig. \ref{fig:captureplot}.  We find that Callisto can be left stranded at its current position for $M_{\rm CPD} < 10^{-6.2} M_{\odot}$.  Above this mass limit, Callisto migrates inward too rapidly and cannot be stranded. The CPD mass that produces the largest region in the initial condition parameter space to reproduce the exact current position of Callisto is $M_{\rm CPD} = 10^{-6.6} M_{\odot}$ (Fig. \ref{fig:captureplot} lower right panel). In this case, Callisto is required to have achieved $90\%$ of its final mass between 26.5 and 30.8 $R_{\rm J}$ and could have migrated at most $\sim$5 $R_{\rm J}$ inward.  For CPDs with mass below $10^{-7} M_{\rm \odot}$ CPD,  the photoevaporation timescale is much shorter than the migration timescale, such that Callisto must have achieved its final mass very near its present-day location. 

It is possible to push the initial position of Callisto out to \mbox{40 $R_{\rm J}$} for a CPD with mass $10^{-6.2} M_{\odot}$. However, the resulting combination of the initial position of Callisto and the outer radius of the CPD that places Callisto near \mbox{26 $R_{\rm J}$} becomes highly constrained (less than $0.1\%$ of the uniformly sampled initial condition space) and it is far more likely to enter the resonance.  Figure  \ref{fig:captureplot} illustrates that Callisto can be placed at its current position for any possible CPD initial outer radius within the considered range, while the range of initial Callisto positions is more tightly constrained.  
Thus far, we have only considered the case of an abrupt exponential decline in mass accretion over \mbox{$\sim10^3$ yr}.  Scenarios wherein the CPD steadily declines in both mass and steady-state truncation radius as accretion slows over several $10^4-10^5$ years allow for Callisto to have grown to its final mass at larger separations and still been captured at \mbox{26 $R_{\rm J}$}, but the steady-state truncation radius for $99\%$ of our CPDs with $\dot M < 10^{-12} M_{\odot}$ yr$^{-1}$ is smaller than 215 $R_{\rm J}$, and in $50\%$ of cases it is smaller than \mbox{35 $ R_{\rm J}$}, placing an upper boundary on the initial position of Callisto.  We therefore consider it unlikely that Callisto formed farther out than 40 $R_{\rm J}$. 

The undifferentiated internal structure of Callisto sets a constraint on its accretion timescale of $> 10^5$ yr \citep{1997Natur.387..264A,ANDERSON2001157, 2003Icar..163..198M,2003Icar..163..232M}.    We note that the minimum accretion rate we considered is able to supply enough mass to the CPD to build Callisto within \mbox{$5\times10^4$ years} in the case of perfect accretion efficiency. Depending on the background FUV field strength, we find that a CPD with an accretion rate $\dot M = 5\times10^{-13} M_{\odot}$ yr$^{-1}$ (slow enough to form Callisto without melting it) is likely to have a truncation radius within the present-day orbit of Callisto for $G_0 = 3000$. The photoevaporative stranding of Callisto can still be consistent with its internal structure for FUV field strengths of $G_0 \lessapprox 10^{3.1}$.

\section{Conclusions}

We have investigated the radiation field to which the CPD of Jupiter may have been exposed during the formation of the Galilean satellites.  Our key findings can be summarized as follows:

\begin{enumerate}

\item Jupiter analogs open gaps that are sufficiently deep to expose their CPDs to interstellar FUV radiation for PPD $\alpha$-viscosities of 10$^{-5}-10^{-3}$. The radiation field of a solar birth cluster analog is sufficient to photoevaporatively truncate a Jovian CPD independent of the steady-state mass of this CPD.  
\vspace{1ex}

\item For accretion rates $\dot M = 10^{-12} M_{\odot}$ yr$^{-1}$ , a Jovian CPD can become truncated to within the current orbit of Callisto \mbox{($\sim 0.04$ $r_{\rm H}$)} for 50\% of the cluster stars, explaining the lack of regular satellites beyond the orbit of Callisto.   The mean truncation radius found in our ensemble of cluster stars is proportional to the CPD accretion rate by \mbox{$r_{\rm trunc} \propto \dot M ^{0.4}$.} 
\vspace{1ex}

\item We find that rapid external photoevaporation of the Jovian CPD following the end of accretion can act to stop the inward gas-driven migration of Callisto and prevent it from piling up into the mean-motion resonance of the inner three Galilean satellites.  The mass of the CPD in this scenario would be constrained to $\leq 10^{-6.2} M_{\odot}$, and Callisto must have achieved at least $90\%$ of its final mass at a semimajor axis between 27-40 $R_{\rm J}$. 


\vspace{1ex}

\item We place an upper limit on the strength of the background FUV radiation field $G_0 \leq 10^{3.1}$ during the formation stage of Callisto.  

\end{enumerate}

\begin{acknowledgements}
The research of N.O., I.K., and  Ch.R. is supported by grants from the Netherlands Organization for Scientific Research (NWO, grant number 614.001.552) and the Netherlands Research School for Astronomy (NOVA). "This research has made use of NASA's Astrophysics Data System Bibliographic Services.  This research made use of Astropy,\footnote{http://www.astropy.org} a community-developed core Python package for Astronomy \citep{astropy:2013, astropy:2018}. The authors would like to thank J. Tjoa, S. Van Mierlo, M. Rovira Navarro, T. Steinke, and J. Szulagyi for their insightful advice and helpful discussion. The authors would also like to thank the anonymous referee for the suggestions and corrections that contributed to improving and clarifying this work.
\end{acknowledgements}

%

\bibliographystyle{aa} 
\bibliography{refs.bib} 

%

\begin{appendix} 

\section{Disk model temperature and density}

\begin{figure}[ht]
    \begin{subfigure}{0.44\textwidth}
        \includegraphics[width=\textwidth]{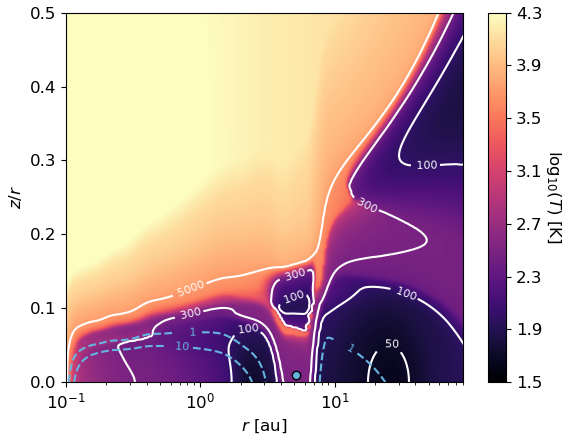}
        \caption{Gas temperature of the solar protoplanetary disk.}
        \label{fig:mmsn_tgas}
    \end{subfigure}
    \begin{subfigure}{0.44\textwidth}
        \includegraphics[width=\textwidth]{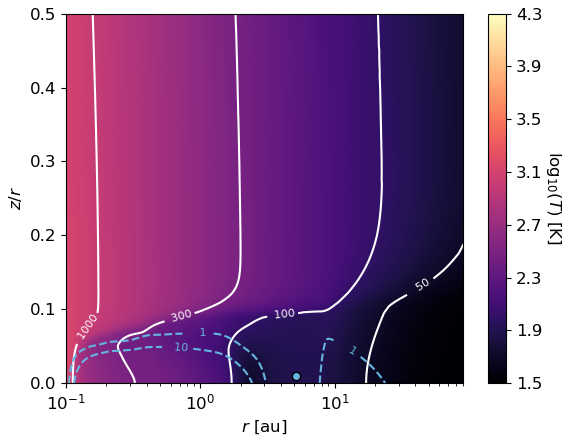}
        \caption{Dust temperature of the solar protoplanetary disk.}
        \label{fig:mmsn_tdust}
    \end{subfigure}
    \begin{subfigure}{0.44\textwidth}
        \includegraphics[width=\textwidth]{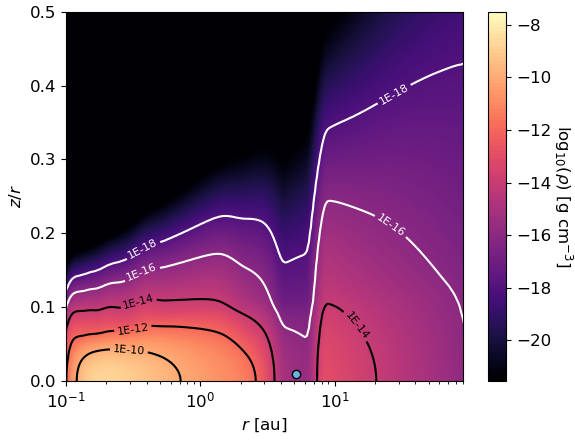}
        \caption{Gas density of the solar protoplanetary disk.}
        \label{fig:mmsn_density}
    \end{subfigure}

    \caption{Temperature and density structure of the solar PPD  \textsc{ProDiMo} model for $t = 5$~Myr, $\alpha = 10^{-4}$.  The dashed blue lines trace surfaces with a visual extinction $A_{\rm V} = 1, 10$, defined by the minimum of either the vertical or radial extinction at a given point.  The blue dot indicates the radial location of Jupiter.}\label{fig:mmsn_all}
\end{figure}

\begin{figure}[ht]
\vspace{7ex}
    \begin{subfigure}{0.44\textwidth}
        \includegraphics[width=\textwidth]{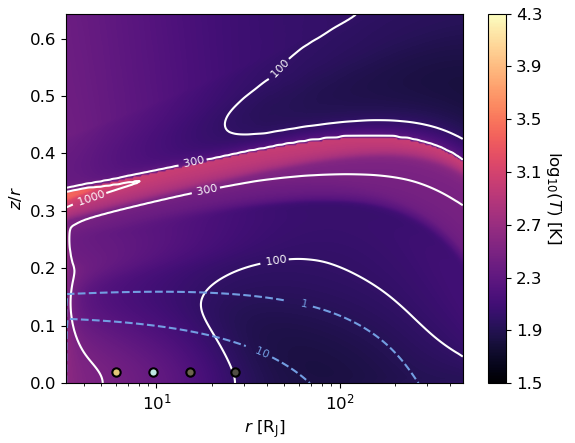}
        \caption{Gas temperature of the CPD.}
        \label{fig:cpd_tgas}
    \end{subfigure}
    \begin{subfigure}{0.44\textwidth}
        \includegraphics[width=\textwidth]{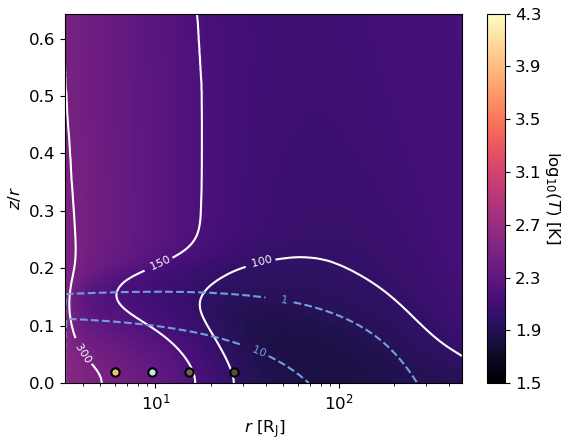}
        \caption{Dust tepmerature of the CPD.}
        \label{fig:cpd_tdust}
    \end{subfigure}
    \begin{subfigure}{0.44\textwidth}
        \includegraphics[width=\textwidth]{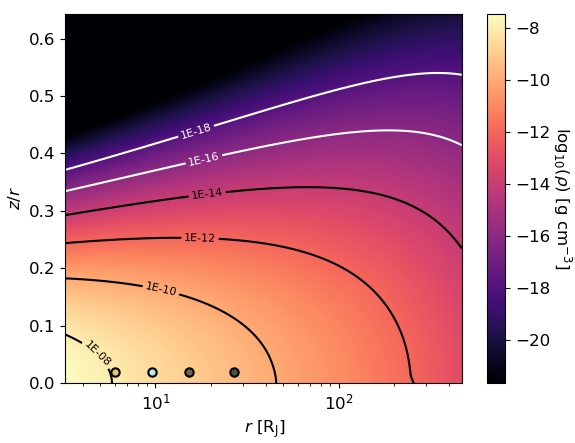}
        \caption{Gas density of the CPD.}
        \label{fig:cpd_density}
    \end{subfigure}
    \caption{Temperature and density structures of the Jovian CPD \textsc{ProDiMo} model of $M =  10^{-8}$ $M_{\odot}$ and $\dot M =  10^{-11}$ $M_{\odot}$ yr$^{-1}$.  The dashed blue lines trace surfaces with a visual extinction $A_{\rm V} = 1$ and 10, defined as the minimum of either the vertical or radial extinction at a given point.  The scatter points indicate the radial location of the Galilean satellites (from left to right: Io, Europa, Ganymede, and Callisto).}\label{fig:cpd_all}
\end{figure}

\end{appendix}

\end{document}